# An Analisys of Application Logs with Splunk : developing an App for the synthetic analysis of data and security incidents.

ROBERTO BRUZZESE[1]

[1]Freelancer, Rome, Lazio, Italy
bruzzese.953247@studenti.uniroma1.it

## ABSTRACT

*The present work aims to enhance the application logs of an hypothetical infrastructure platform, and to build an App that displays the synthetic data about performance, anomalies and security incidents synthesized in the form of a Dashboard. The reference architecture, with multiple applications and multiple HW distribution, implementing a Service Oriented Architecture, is a real case of which the details have been abstracted because we want to extend the concept to all architectures with similar characteristics.*

.

## KEYWORDS

*Log Analysis, Data Analysis, Machine Learning, Splunk, Security Incidents*

## 1. INTRODUCTION

In this chapter it is described the work done in the context of the experimentation on Splunk tool in analyzing Logs. The work starts from the initial point, the search for logs, and continues with the problems to face for manipulation of raw data, the control rules for detecting web attacks, ending with a Machine Learning and statistical approach to the determination of anomalies.

## 1.1. Description of the starting point

The starting point of my trip is a collection of files, the logs, within a medium-high complexity application infrastructure. My aim is to make these objects becoming a resource, more properly an asset of the organization.

The logs inside the Application Infrastructure (for the description of the architecture see the Figure 27) are many and often scattered in the meanders of the machines and their directories, and are often not known at company level except by individual users.

They are often used by individual developers who use them for troubleshooting purposes, or by system administrators who use them for management purposes, or by security officers for control purposes.

However, there is a lack of a unified vision of the resource, a lack of cross analysis between the various logs and of an in-depth look at the information they contain.

It often happens that these files, sometimes very voluminous, are not even read, before being forgotten or completely deleted. For this reason, it has origin the present work with the need to try to enhance, through an approach of "exploration and discovery", of what is hidden in those logs. Then my activity consisted in what can be called "log mining ".

As being perfectly aware that logs, intended as a support to contain a record of the activities carried out in an information system, can come from very different sources (firewalls, servers, databases, networks, applications, etc..) and that my main purpose is to explore only the contents of logs of application origin.

## 1.2. Fragmentary knowledge of available logs

Often every single developer knows the location of their own logs. You have to look for a little bit before you find what you are looking for, if, as in my case, you play a cross-cutting role in applications. In an Infrastructure like the one under consideration (which is based on the J2EE platform and which is built with a SOA approach , using IBM Websphere Application Server) , it is well known in advance the position and the format of the Application Server logs, since they are standardized.

As far as the logs of the single applications of which the infrastructure is composed are concerned, it could be searched in the directories of the respective applications, where the logs are produced.

These logs are the ones on which the attention is focused since they are of direct competence of a software development unit in which this work was carried out.

The evaluation of the format of the individual logs made in the first analysis let me believe my work could pursue several objectives:

(1) to proceed to an automatic extraction of targeted information
(2) measure quantities, evaluate trends
(3) detecting errors, reporting
(4) exploring relations between information

## 1.3. It should be decided to adopt a specific or general solution ? Make or buy ? Open source or commercial ?

A possible choice was to build a log analysis [1] solution starting from open source resources, freely available, and of limited cost.

In particular, the "grep "commands could be used. This command is a Linux or Unix utility that searches file input patterns or regular expressions.

This type of command could allow us to focus on a single problem at a time and examine the results in detail.

However it seemed to be more targeted for small cases or for analysis that are carried out with a high degree of manual skills and also have a low "friendly" type degree of usability. It also seemed that the integration into an infrastructure was restrictive and cumbersome, and complex, above all in the case of multiple types of different logs.

In addition, there are the following factors that are detrimental to this "grep utility" tool:

- It is suitable when you know perfectly what you want to achieve, and you focus on very specific text, on the whole file.

- You should not overlook the time it takes to search for large files.

- It seems often endless as you search all over the file (which can be terabytes!).

- It is not easy for correlating searches on multiple log files at the same time, or even on different devices (it may be useful to extend the search on the Web server of the Web Farm, or on the HTTP server, or on a Firewall).

The latter is an important aspect in log management because, in a multi-application and multi-distributed infrastructure , the application log is produced on the machine on which the application is hosted and executed, and in the case of cross analysis of logs from different sources, it is necessary to centralize it as the next Figure 1:

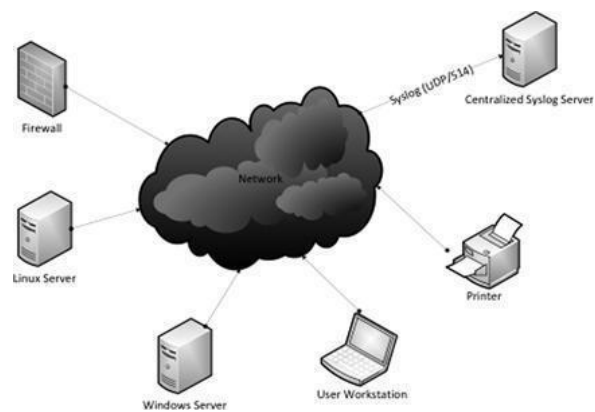

**Figure 1:** Centralization of Logs.

Splunk [2] commercial tool immediately seemed to me the most viable solution, providing this company the Enterprise distributions, freely downloadable, with licenses valid for 60 days, or, alternatively, Developer licenses, valid for 6 months.

This tool offers a series of features that reduce the time needed to analyze the logs on a periodic basis, as there are activatable files/directories monitors, which update the analysis. It also makes available the generation of alerts at the occurrence of predefined events, the generation of reports that allow corporate compliance, the centralization of the logs to be analyzed on which to make further actions or investigations. It should not be underestimated that several add-ons do exist, and also apps and APIs that allow Splunk to be integrated with the surrounding infrastructure.

Its use allows the easy centralization of log management, through different instances of log collector (Forwarders) , or log analyzer (Head Searches). It is possible to analyze multiple sources simultaneously, without affecting performance, as Splunk is based on the concept of data indexing (Indexer) that optimizes search times.

It is also easy to report and display in a Dashboard the summary data of the search results, thanks to a rich availability of graphs, which can be used from a web interface or even via a command line or web services.

## 2. SPLUNK FUNCTIONAL ARCHITECTURE

### 2.1. Splunk Functional Architecture

It is possible to install the version of Splunk Enterprise on a single instance of a machine, in such a way to have on a single site all the peculiarities of Splunk, namely "data input", "indexing", "search". In the case of more complex organizations, the typical architecture of Splunk is the type shown in the following Figure 2. The architecture is functional to the organization in which you want to install Splunk.

### 2.1.1 The Forwarders

Forwarders are 20 mb software distributions which can be installed directly on the site where the data need to be collected. They only monitor files or directories and send the collected data to Splunk instances (Enterprise type) that provide indexing of data.

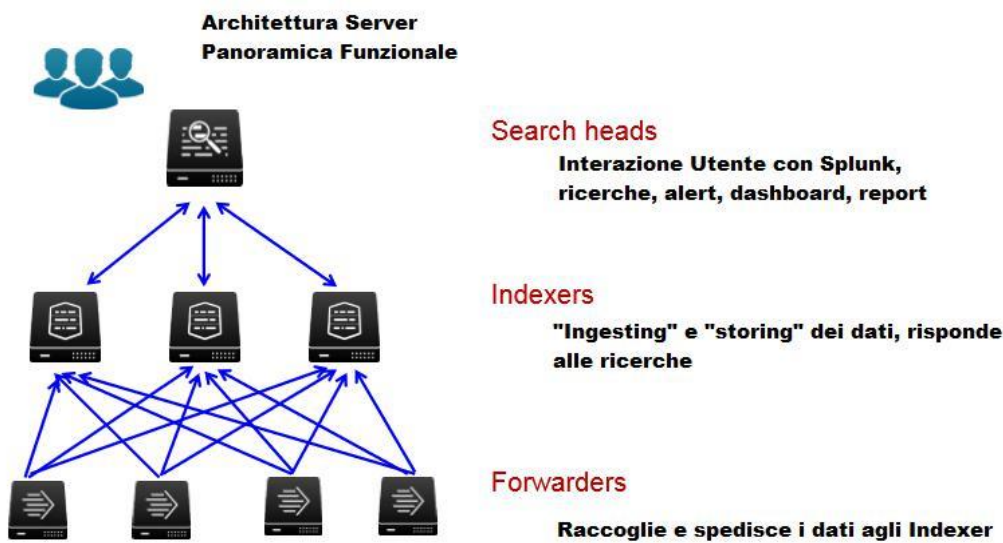

**Figure 2:** Splunk Functional Architecture Overview.

### 2.1.2. The Indexers

While the number of Forwarders is equivalent to the number of machines on which there are logs that need to be analyzed, the number of Indexers is calculated on the basis of a capacity plan depending on the size of the logs that are estimated to be handled. It is necessary to have approximately 100GB for each Indexer, in a configuration that meets the hardware requirements provided by Splunk. On the Indexers takes place data parsing and data indexing.

### 2.1.3. The Search Heads

As far as concerns the Search Heads, which are the input end points for user interactions, their number depends on the number of Users expected for that configuration. It should be noted that a Search Head User means both an interactive user and an REST application or also a web service user.

### 2.2. Initial Architecture

The architecture implemented in this project, is described in the following Figure 3.

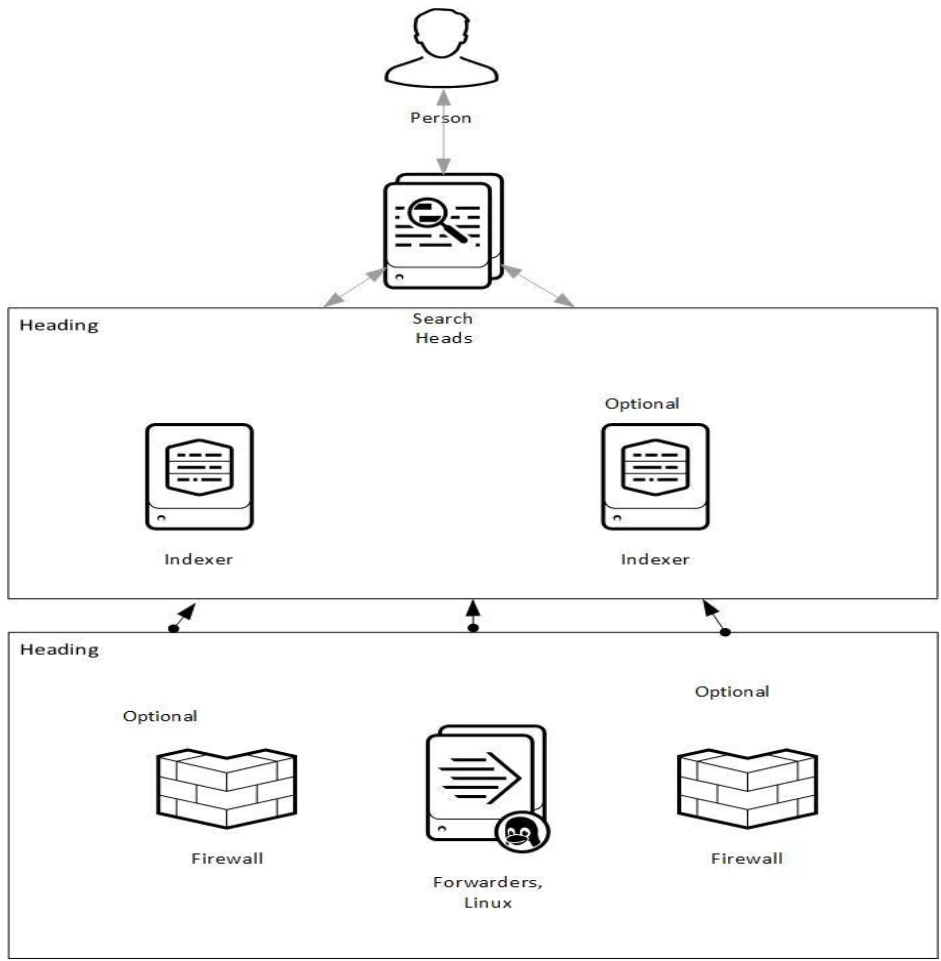

**Figure 3:** Initial Architecture.

This architecture has been realized using Virtual Machines created on the Aruba Cloud VPS. The instances were created by installing Splunk and transferring the logs which were withdrawed from the production environment of the Infrastructure. The instances of Splunk which were created are reported below :

**Universal Forwarder** : 1 of GB RAM, 1 single core CPU, 20 GB SDD, Linux system

**Indexer** : 2 GB RAM, 1 CPU single core, 40 GB SDD, Linux system

**Search Head** : 2GB RAM, 1 single core CPU, 40GB SDD, Linux system

Forwarder machines communicate with Indexer machines via a TCP connection on port 9997, through which they send data to the Indexer.

This minimum configuration is adequate to guarantee acceptable performance with a maximum of 5-10 logs, of about 300 mb size each and 1-2 million events.

## 2.3. HW System Requirements

The hardware requirements reported in Splunk Documentation necessary to be an adequate configuration to ensure efficient log management and searching are listed here below :

### Basic Level - Single Instance Host

Intel x86 64-bit chip architecture 12 CPU cores a 2Ghz (at least 12GB RAM Standard 1Gb Ethernet NIC 800 IOPS (on average) Performance of Indexing Max : 20MB/sec Performance of Search Max : 50.000 events/sec

### Medium Level - Distributed Multi Hosts

Instance Intel x86 64-bit chip architecture 12-16 CPU cores 2Ghz (at least) 12GB RAM Standard 1Gb Ethernet NIC 800 IOPS (on average)

### High Level - Distributed Multi Hosts

Instance Intel 64-bit chip architecture 48 CPU cores at 2GHz (at least) 128GB RAM SSD 1Gb Ethernet NIC plus NIC optional SAN

### 2.4. Splunk processing: data pipeline

The initial architecture distinguishes the functional activities of Splunk, which take place as a data processing chain. Regardless of whether a single-instance or multi-instance architecture is adopted, the "data pipeline" follows three distinct stages:

1. Input: where the data is preloaded, divided into 64K blocks and annotated with the metadata: host, source, sourcetype. At this stage, the concept of event does not yet exist.

2. Parsing: In this phase Splunk segments the data into events, sets the timestamps, notes the metadata it obtains from the previous input phase, possibly transforms the event data, according to the transformation rules (regex).

3. Indexing: the events obtained by parsing are written to the disk in compressed and indexed form.

4. Search: this phase concerns the user's access to the indexed data (in the form of search, reports, alerts, dashboards).

In the following Figure 4 is showed the data pipeline process :

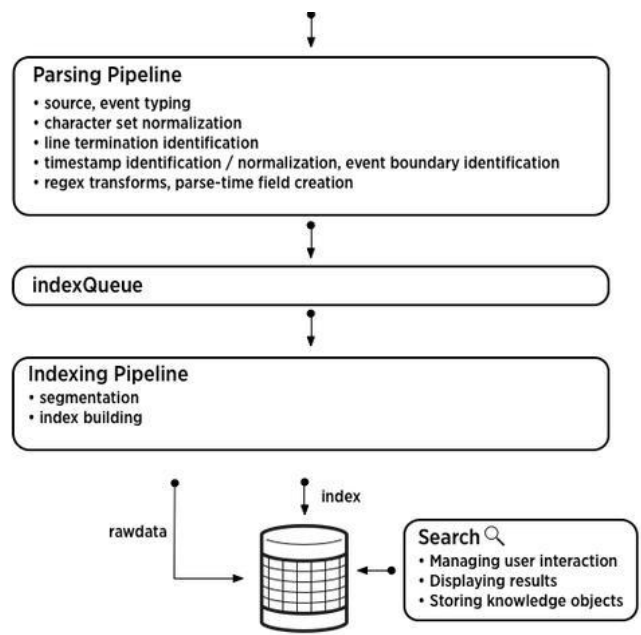

**Figure 4:** Data Pipeline.

Although in the present project there is a single source of data, it can be the most varied as Splunk is predisposed to all sources and formats (Figure 5).

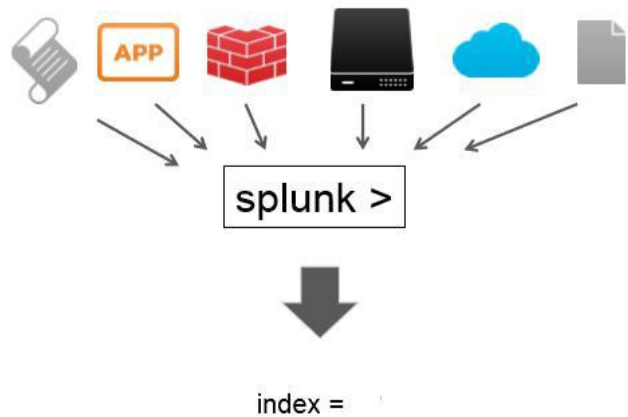

**Figure 5:** Heterogeneity of Data Sources.

## 3. INGESTION, EXTRACTION, ANALYSIS

### 3.1. Analysis of the Infrastructure Logs

In this project were analyzed sample portions of the Infrastructure Application Logs. In particular those related to a couple of applications.

Three main Indexes (for the sake of simplicity we will denominate them indexA, indexB and IndexC) have been created, on which all the searches have been carried out.

The set of search queries submitted to the system are still general and can be repeated in multiple contexts. In order to increase the generalization, the ease of use and the reusability of the search queries, the corresponding macros have been stored (although not reported in this paper).

After loading, parsing, and indexing the data, it was necessary to pre-treat the data, in order to extract and better define the fields that are used in the subsequent stages of search.

### 3.2. Data Loading and Extraction

Data Loading into Splunk is an important step for the subsequent analysis and visualization phases. If the data are not correctly analyzed, time-stamped, and divided into events, it will be difficult in the next stages to carry out a 'proper analysis of the data and get insights into them'.

Since the data may come from many sources as web servers, application servers, operating systems, network security, services or applications (although in our case they will be exclusively of application origin) it is still important that the format makes all the subsequent analytical phases easily searchable.

In general, Splunk is equipped with internal facilities that recognize the most common formats from various sources (Figure 6) and then apply the preconfigured settings already stored. If Splunk recognizes the data source, it will apply the settings and definitions for that data source. The data formats can be divided into three types: **structured**, **semi-structured**, and **unstructured ones**.

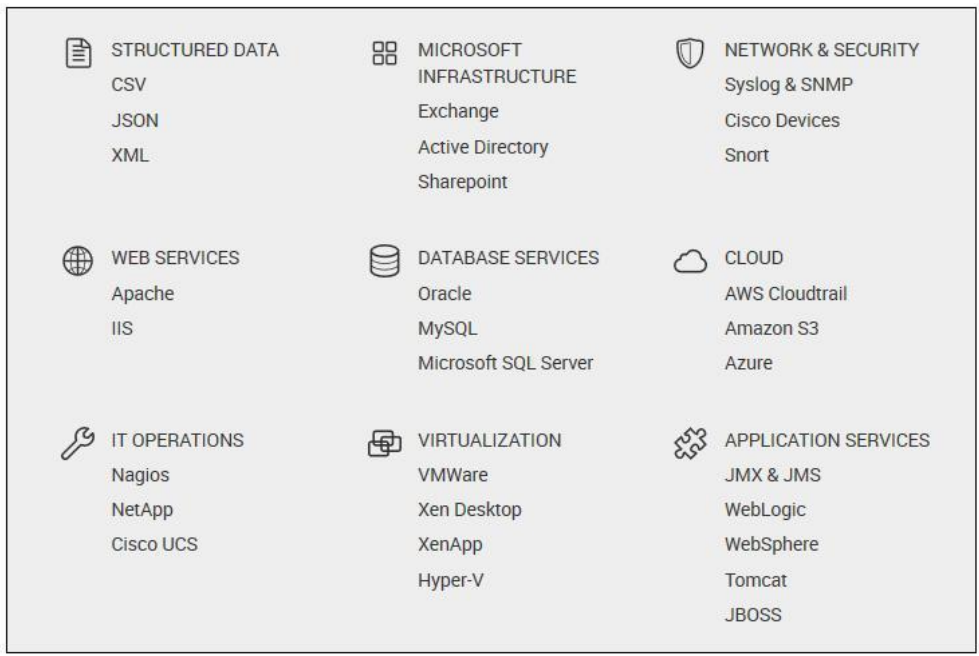

**Figure 6**: Preconfigured Splunk Data Formats.

If structured formats occurs maybe additional settings need to be specified for the custom format in question. This can be done in the field extraction phase.

### 3.2.1 Application Logging

In the case of the application infrastructure under consideration, the logs considered are basically related to applications developed in-house. In this case, logging is delegated to developers for troubleshooting reasons, or for determining a certain level of quality service or determine system availability (through measurements of systems interruptions).

Most of the time they are just the developers who understand the meaning of the application logs as they are ultimately the only users to determine software bugs or make a fixing in case of malfunctions.

In the kind of infrastructure under consideration, which is a medium-high complexity system, with about 20 machines cooperating in the architecture, the resolution of application problems is a complex and time-consuming activity.

The adoption of a tool like Splunk, thanks to the use of queries and reports, alerts and messaging, can be of fundamental help to maintenance and operational activities.

This requires an activity of extraction of the fields, in order to fully analyze data, and thus making data to become an asset for the entire organization.

In fact, because it makes sense to believe that 'garbage in ... garbage out', it cannot be expected to extract value if into Splunk is introduced only information of little value. In this sense we should follow as much as possible an approach in which there is a minimum gap between the development and operation of the system. Basically you have to try to get a log that is 'machine friendly' and 'human readable'.

### 3.2.2 The Loggers

At the application level, the developer uses loggers (i.e. java log4j) that allow to write messages in multithreaded mode and in a typical standard format. In this way, if there is an exception, some peculiar aspects can be checked, while others are established and delegated to the Java framework.

In other words, the logger keeps track of what happened in the program, so if a crash or bug occurs, you can trace the cause.

We can say that the events produced by the application loggers are in the form of '**structured data**' while those that come from the proprietary applications developed ad hoc generally have a form of '**semi-structured**' data and in other cases of '**unstructured data**'. The latter may vary, and may also extend over several lines, while the event they report may be in a format without precise separators, and without the obligation of a start or an end clearly identified or reported explicitly.

The logs here reported and examined, relevant to the Infrastructure, are mostly structured data, although in some sections of the logs there may appear unstructured ones, as shown in Figure 7 :

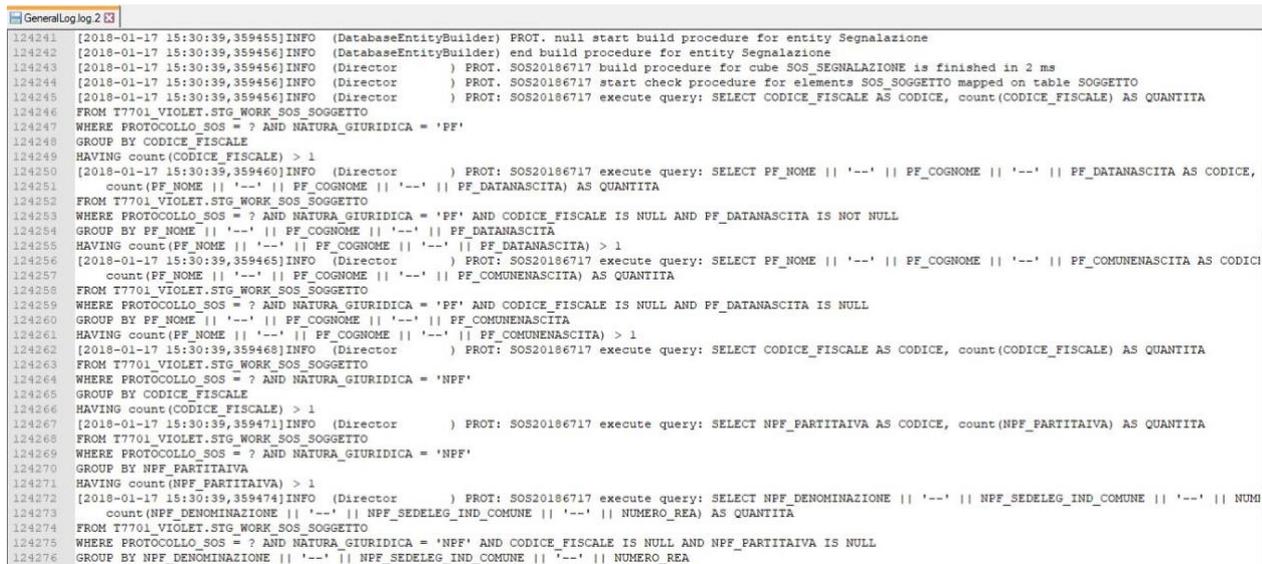

**Figure 7**: Format of the Log indexA.

This latter type of log is much more complicated because all knowledge must be extracted manually and it requires collaboration and information exchange between the Splunk engineer and the programmer or vendor. In such cases, if Splunk is allowed to operate automatically, it will try to extract stuff that it wrongly believes to be data, and the whole process will not bring added value in the following phases of analysis and reporting.

In such type of logs, in order to be able to carry out a correct extraction of significant fields, it must taken into account the **'event breaking'**.

To make sure that we are loading or extracting correctly, we must put order in the initial chaos and this usually begins by separating an event from the next one, by means of a line or a character. There are many ways in which you can separate events from each other (see Splunk /Admin/Propsconf Documentation).

### 3.2.3. Data Extraction

The step mentioned above is called **'event extraction'**. It consists of identifying the line or group of lines that constitutes the information relating to an action. The **'extraction of a field'** consists in identifying precisely what can be considered a field, for its sequential position in a list of values, assuming the field is not already in the form of a key=value. This is a process for the that Splunk administrator, which creates value in the data sets.

### 3.2.4. Metadata fields

It should be noted that there are three most important default fields: host, source, and sourcetype which are also called metadata. They describe where the event originated.

**'host'** – it indicates the name of the host, the device, the IP address, or the domain name of the network from which the event originates.

**'source'** – it indicates the name of the File, stream, or any other input from which the event originates. In the case that the data comes from directories and files specified by the configuration files, the source value is the path, e.g. /archive/server1/var/log/messages.0 or /var/log/. If the data source was based on data from a network, such as a protocol or port, it would be UDP:514.

**'sourcetype'** – this field stands for the format of the data of the source from which it originates. This value determines how the data will be formatted. Other default fields include date and time fields, which add more search 'granularity' to event timestamps. Finally Splunk Enterprise adds other default fields classified as 'internal' fields.

## 4. SEARCH ON APPLICATION LOGS

### 4.1 Examples of searches on indexes

Once the meaningful fields have been extracted, using the interactive extractor and the regular expressions, the search query can be run. This is possible using SPL (Search Processing Language) and after the extraction of of the relevant fields. Here below will be reported the main searches compiled for the present work.

### 4.1.1 Query 1 : SQL Query Times

This search query asks the system, to list the first 10 time intervals spent in running the sql queries of the application, sorted by percentage distribution, from the first, most frequent, to the last. As it can be seen from the result of the query in Figure 8 , 47% of SQL queries last 1 second; about 29% last 2 seconds.

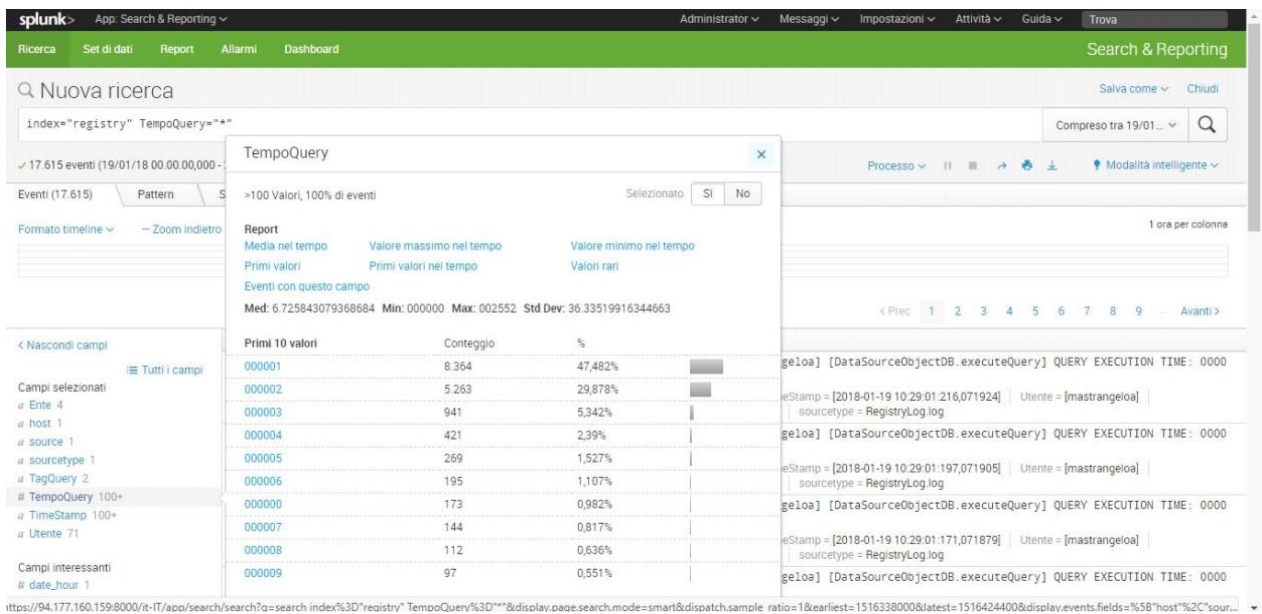

**Figure 8:** results of Query 1

### 4.1.2 Query 2 : SQL Query Times (actualized)

In this query, carried out on the log, we asked the system to make a 'transaction', that is to group events having in common certain characteristics. In particular, it was asked to display all the queries on the axis of the abscissae, and the relative times taken to run that query, on the axis of the ordinates. By passing the mouse on the single line of the histogram, the query in question can be displayed. You can see, in the Figure 9, for the purpose of determining possible anomalies, a query whose time is more than 1500 ms.

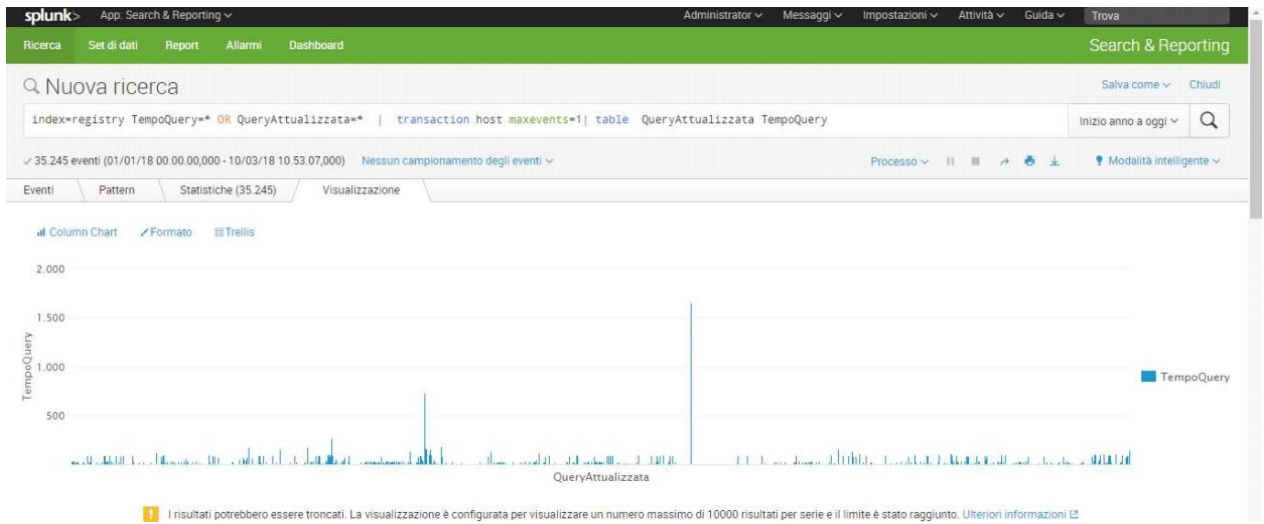

**Figure 9:** Results of Query 2

### 4.1.3 Query 3 : SQL Query Times (anomalies)

In this query shown in the Figure 10, the maximum value recorded on all queries ran on the system is asked. This value is enough high to be considered an 'outlier'.

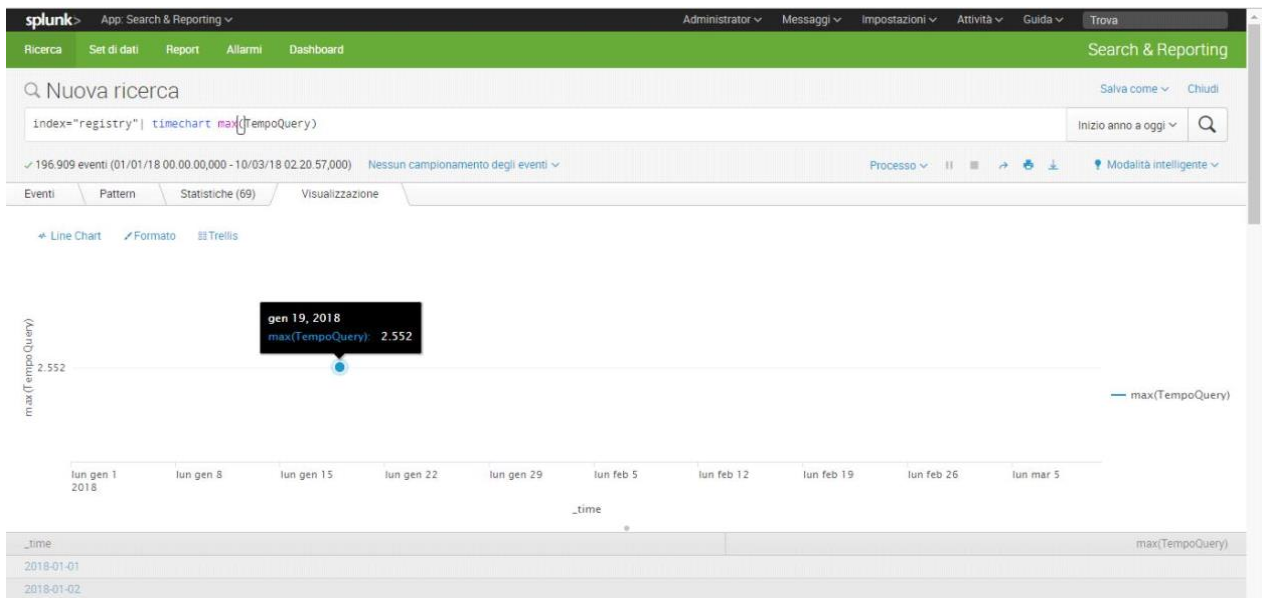

**Figure 10:** Results of Query 3

### 4.1.4 Query 4: Incomplete Transactions on Services

In the following search query, carried out on the index log, it was asked to display all incomplete transactions, i.e. all those that have a service login that does not match a service logout, and/or viceversa. In the pie chart in the Figure 11 , there are 1271 total events found, distributed on two dates, 17/1/2018 and 18/1/2018.

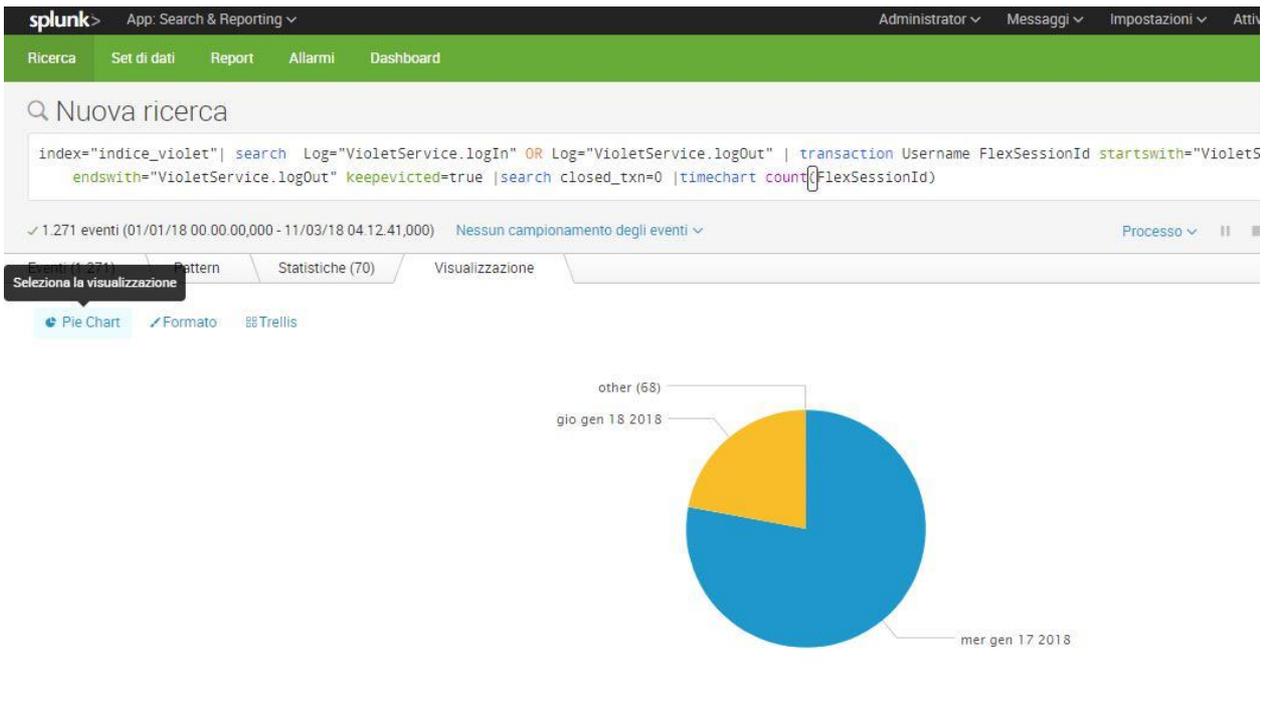

**Figure 11:** Results of Query 4

### 4.1.5 Query 5 : Measure of Application Load : 'transaction'

In this query (Figure 12), an estimation of the application load is inferred, by assuming that the "Log" field, gives us an approximation of this measure. Since we can't directly measure CPU cycles, the number of service operations, indicated by "Log" field, divided by time slot, is likely one of the possible load measurements.

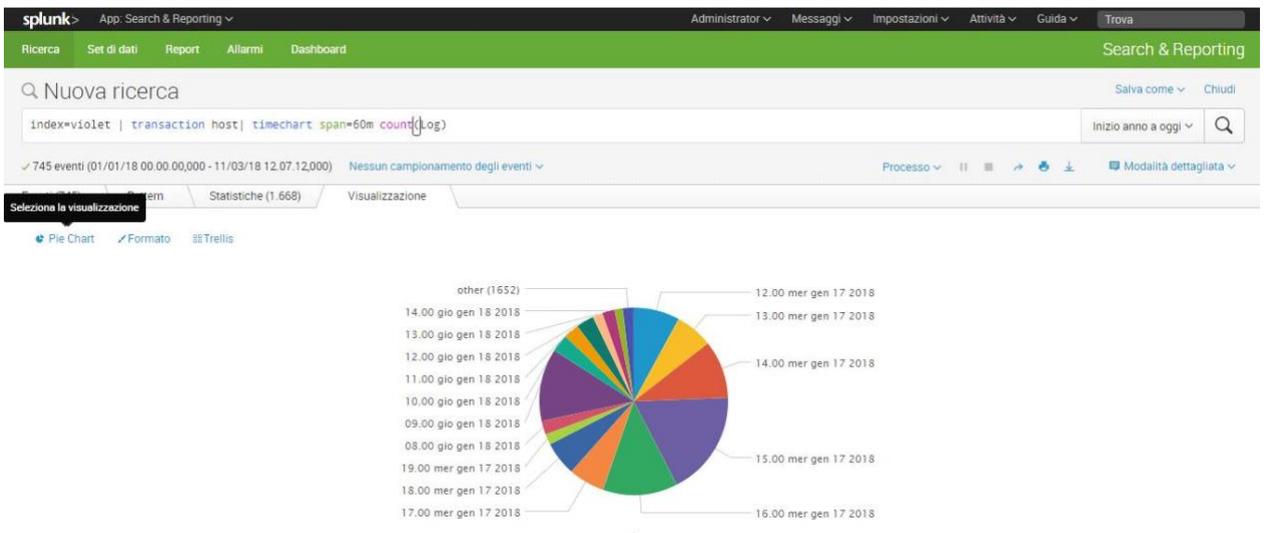

**Figure 12**: Results of Query 5

### 4.1.6 Query 6 : Application Load Measurement : 'System pauses'

In the following query carried out on the log (Figure 13), it was asked to highlight all system pauses longer than 2 seconds, and report them with respect to the time on the abscissae axis. The pauses are deduced from the event timestamp. The interpretation given to the result of the query is that there are periodic instants of time, coinciding with the hours or half hours, in which there are breaks greater than 2 seconds. It can be supposed that these application pauses may coincide with higher-level tasks, which the system or application must perform in those deadlines.

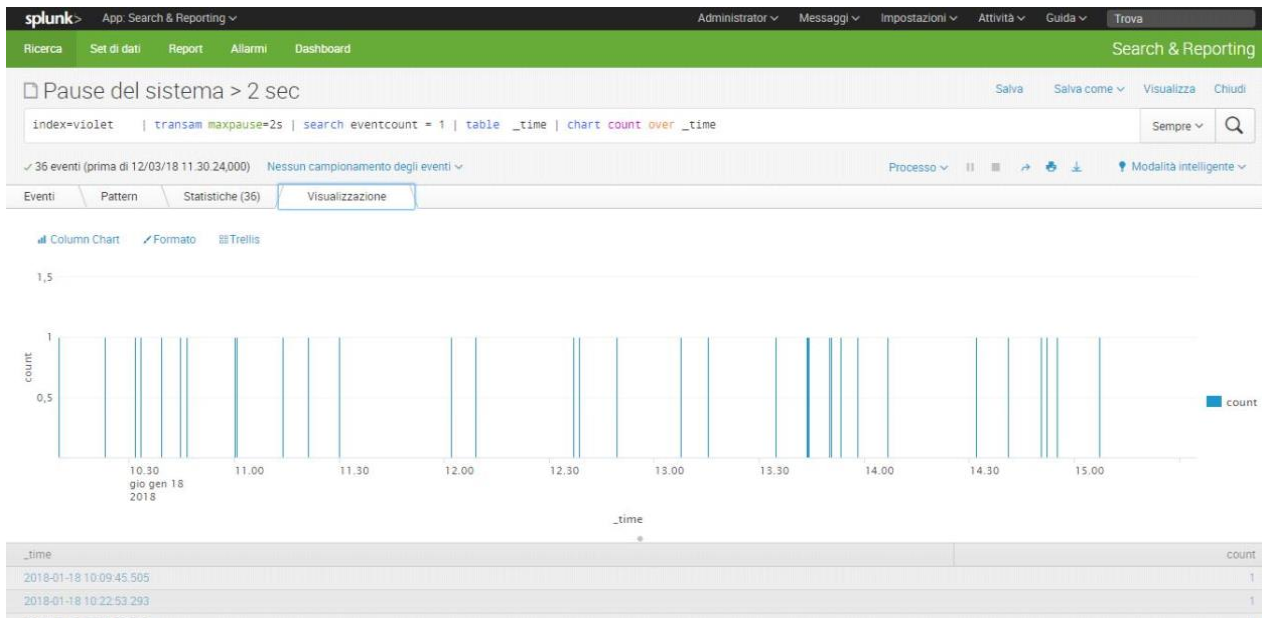

**Figure 13:** Results of Query 6

### 4.1.7 Query 7 : Application Load Measurement : 'events concentration'

This query also can be considered a measure of the application effort of our system, as it is a distribution of the time intervals between two successive events. As you can see in Figure 14 , there are more events at a distance of 0 ms, in respect to those at 0.1 ms, and to those at 0.2 ms time distance. This fact highlights a big commitment of the system also, and probably, in a multithreading mode.

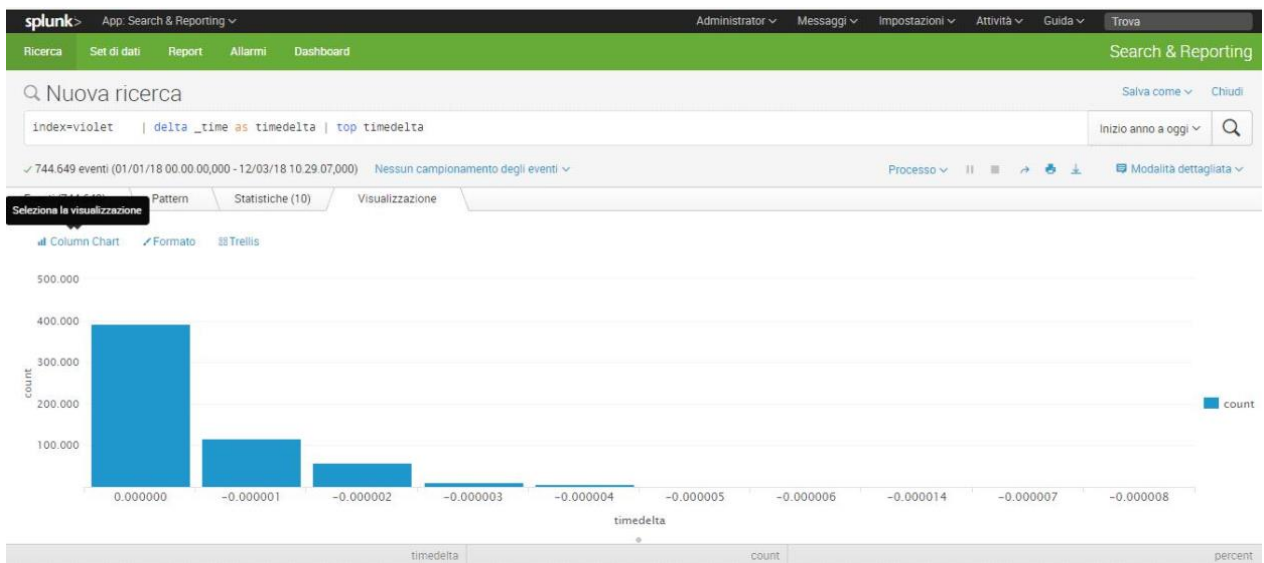

**Figure 14:** Results of Query 7

### 4.1.8 Query 8: Broken Authentication and Session Management Control

This query is genuinely a security check to ensure against attacks such as 'Broken Authentication and Session Management'. This security check is done in the web server log file and could be useful for both external and internal access verification. There is a type of attack of this type called 'session fixation attack' that occurs by passing on the URL, in the request, the session identifiers preset and invalid. In addition, in the search query, neither the user nor the password should be passed in full, as this constitutes an exposure of sensitive data in the URLs logged into the proxie web or stored in the cache. The security control contains the following control strings:

1) /login\.jsp.*\?.*(userId|password)=./

Which is the login.jsp script followed by ? , user and password

2) /;jsessionid=./

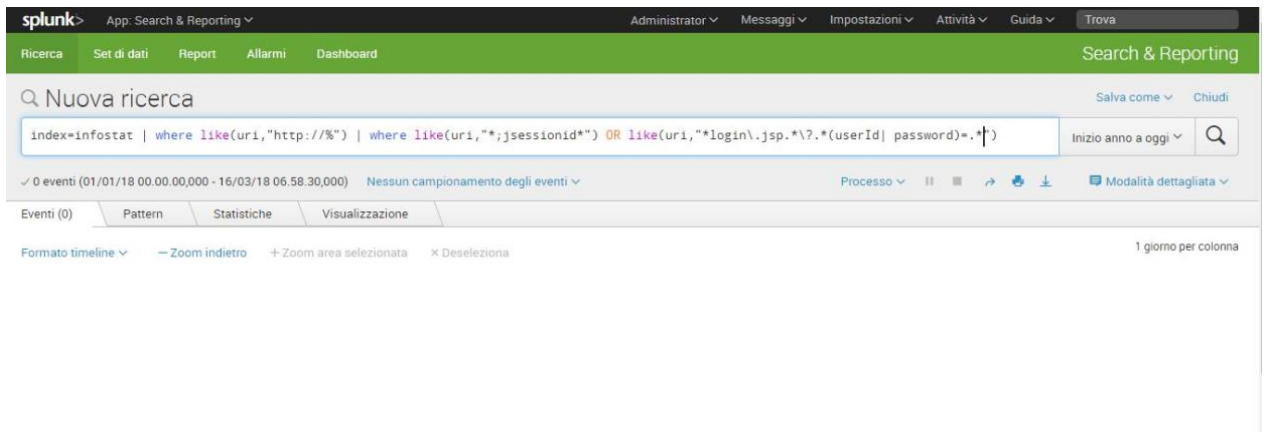

**Figure 15:** Results of Query 8

In the following pages will be reported further control rules against possible attacks, according to the OWASP classification, potentially taking place in the web farm, and recorded on the logs.

## 5. QUERY EXECUTION TIMES AND PERFORMANCE CONSIDERATIONS

Here below are reported some execution times in order to evaluate Splunk performances and its degree of interactivity with the user. Obviously it should be taken into account that the machines have been designed with very low power compared to what is required by the HW requirements. This is because of cost limitation reasons, and also because of the stress test of the system was not in the main objectives of the present work.

Not all search queries ran in this research prototype are the same in terms of power and resource requirements. For example, the commands such as 'cluster', 'transaction', 'correlated', 'transam', as well as the search combined with pipes in sequence, absorb a lot of CPU and therefore increase waiting times, which rarely go above a minute even for the most complex queries.

Queries can be optimized to increase performance. In addition, if called in program mode, they are further optimized as they do not engage the web interface.

To examine the time taken just inspect the process performed for the query, as shown below. For the execution of the query : "ERROR" the inspection of the process reports the following times:

1,018 seconds for the execution of the query :

**Figure 16**: Error Process Inspection

to execute the query: 'CaricoApplicativoLog' the process inspection reports the following times:

101.36 seconds for the execution of the query :

**Figure 17:** CaricoApplicativoLog Process Inspection

# 6. THE DASHBOARD AND THE WEB

## 6.1 The Infrastructure Application Dashboard

A Dashboard has been created (Figure 18) , that summarizes the data obtained from previous search queries, inserting the results in quadrants that rapresent the number of errors or warnings, the performances, the application load, the security alerts. The idea is that these values could eventually be summarized in a single value, the Key Performance Indicator (KPI), which indicates the overall performance of the Infrastructure application system and should be linked to the Service Level Agreements established by the organization, to internal and external users.

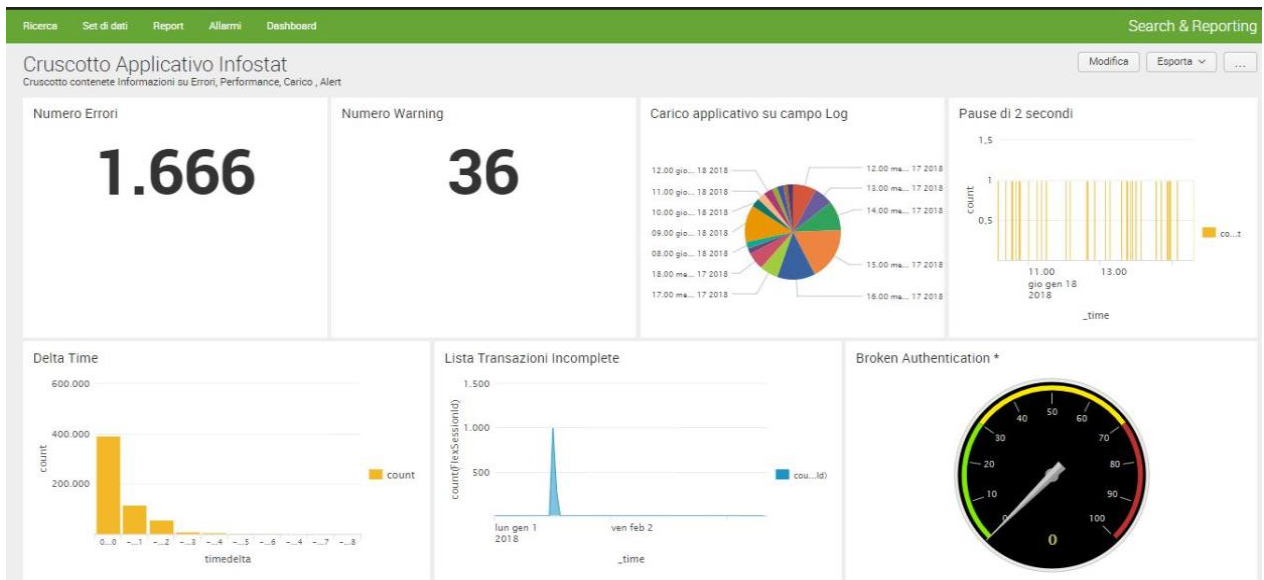

**Figure 18**: Infrastructure Application Dashboard

## 6.2 Bringing Splunk to the Web

The Dashboard created so far is a synthesis of information extracted from the logs of the Infrastructure application. As it has been realized, it suffers from a certain staticity and therefore the aim of the work was to make its use more interactive in order to allow those users who consult it to have more freedom of action and deepening or drilling into its values. It was considered also as an important point the portability aspect and the easy installation on any machine of the infrastructure. For this reason it was decided to make it an app.

An app in Splunk is nothing more than a set of configuration files.

Even if the dashboard is a way of rendering a lot of information in a synthetic, visual and strategic way, without the need to use SPL commands, it is important to give the user the possibility to explore it further.

For these reasons this APP was created which has its own navigation menu and which displays the application dashboard.

The app, set in a package, has been evaluated as meeting the criteria defined by Splunk with the AppInspect control, and therefore ensures its quality and robustness standards.

It can be activated by clicking on the icon of the Splunk apps installed and it has been 'branded' by inserting the logo of the Department of Informatics. By clicking on this icon the application dashboard, shown in the following image, can be accessed (Figure 19)

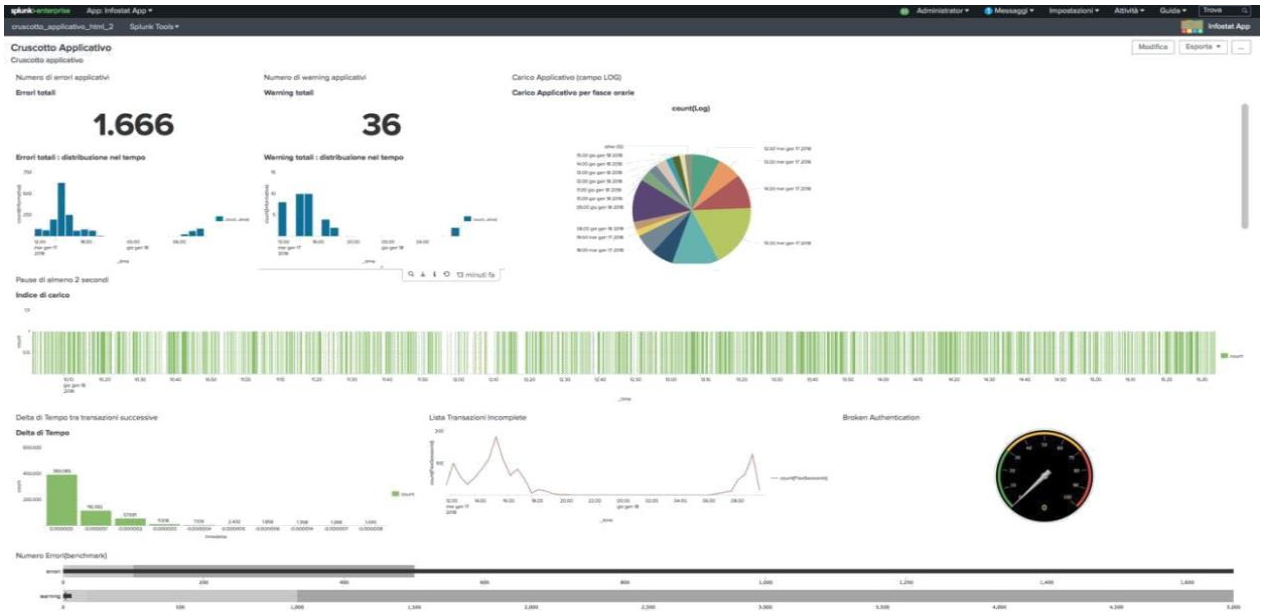

**Figure 19**: Dynamic Application Dashboard

# 7. HACKER'S ATTACK CONTROL RULES

## 7.1 Control rules for the safety of the Infrastructure

The infrastructure architecture we have taken under consideration in our research project is a distributed one, with multiple applications cooperating. It is a Service Oriented Architecture and implements parallelism for the purpose of load balancing and increase of reliability. The Infrastucture is divided into a web farm front, i.e. the component exposed to the Internet, and a server farm, i.e. the internal back-end. The logs analyzed so far are only a very small part of the logs of the Infrastructure. This choice has been made because the logs selected have been considered relevant to the chosen portion of time (two days), and for they were extracted from a single machine, since the attention was limited to a couple of applications, the ones under developers responsibility, who are committed to troubleshooting activities. Only application logs were examined because this was the context in which the author's work takes place, being himself an application developer.

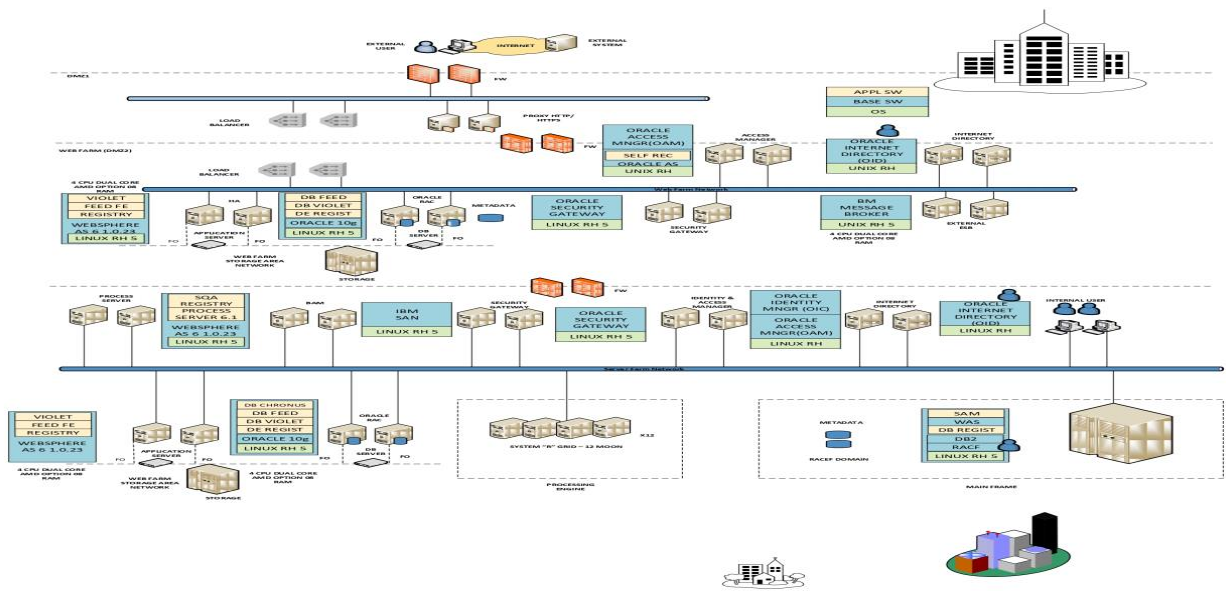

**Figure 20:** Infrastructure Architecture

The total size of the logs analyzed is about 2GB, covering hundreds of millions of events. The analysis which was carried out, mainly revealed anomalies, and strings were also scanned in order to identify fraudulent authentication attempts.

There is a distinction between the log of a front end and the log of a back end, since as the former has information about connections from external clients (Http), the latter mainly operates connections with databases or web services (SOAP). Obviously the logs of the applications in the web farm can allow to identify the attacks coming from outside, while the logs of the applications in the server farm allows to identify attacks form inside.

In the case of attacks from outside, because the field of intervention is level 7 of the OSI architecture (that is the application layer), and because the information used to determine security attacks consists of logs, the rules that will be activated will be based on the 'uri' field, in order to identify the major **security attacks** as cataloged by the organization **OWASP**. These type of attacks, in general, have the purpose of sabotage, or theft of information, or the compromise of the integrity of the data itself. In this case data can be changed and not only stolen. It should be assumed that, in addition to the application logs, should also be scanned the access logs of the Application Server.

### 7.1.2 Cross Site Scripting (XSS) Control Rule

The *XSS attack* can be identified by detecting HTML tags relevant to the execution of scripts in the uri field. Therefore it is a matter of entering a uri field search with the following SPL commands:

'index=indexC | where like(uri,'http:/%') | where like(uri,'*<*>*</*>') | chart count'

this search should identify all suspicious requests like : <script>alert('XSS')</script> or <h1> alert('XSS')</h1>

An alternative method is to explicitly list expressions that can activate active codes or javascript.

The SPL control rule in this case should be :

'index=indexC | where like(uri,'http:/%') | where like(uri,'*javascript*') like(uri,'*vbscript*') OR like(uri,'*applet*') like(uri,'*script*') like(uri,'*frame*') |chart count'

### 7.1.3 Cross Site Request Forgery (CSRF) Control Rule

There are many applications that are vulnerable to CSRF attack. They are those based on the automatic submission of credentials, as session cookies. The CSRF attack, which therefore presupposes an application with vulnerabilities (as an example we suppose it is hosted on the site www.example.com), can consist of putting a post in a forum with a link of the following type:

''

The request will start from the forum site and will be directed to the destination site (trying to download the image size 0).

The attack therefore comes from the forum site, at the implicit request of the attacker, who does not perform the action directly.

To identify this attack in an access log file, is possible through the use of the referer. In the log, in fact, it will be present the referrer, that the client reports to be the site from which the reference or the link to the resource, requested. The one which the GET has started. From the examination of the log relative to an application that suffers this type of attack we can note something that usually does not happen, that means that the URL (or URI) request, that carries out the transaction, and the referrer, are belonging to two different sites. The SPL commands will be as follows :

**'index=indexC | where like(uri,'http:/%') | where IP != referer'**

The referrer value will also be an IP as a result of a lookup table or a workflow action that translates the domain into its IP address.

### 7.1.4 SQL Injection Control Rule

Let's consider one of the most recurrent injections, even if many others can be identified. Injection is made using the characters () or (), which are delimiters for queries and comments, and also the character (#). The SPL commands used to check the attack will be :

**'index=indexC | where like(uri,'http:/%') | where like(uri,'*'*') OR like(uri,'*(userId)=*')' )| OR like(uri,'*(password)='1'='1'-') | chart count'**

A possible attack vector, among many possible, could be as follows: ') or '1'='1. Of course there are endless variations of these values as 1' or 1<2. The constant part is the single apex followed by the term OR.

### 7.1.5 Mischievous File Execution Check Rule

Since applications can allow the user to provide a filename, or part of a filename, this constitutes a vulnerability in case the input is not validated. Therefore, the attacker can manipulate the filename to run a system program or an external URL. For example, the attacker may attempt to upload an executable file, or a file that calls other files to run or parse. The SPL commands to chech this attack will be :

**'index=indexC | where like(uri,'http:/%') | where like(uri,'*.jsp*') OR like(uri,'*.xml*') | chart count'**

## 7.2 Control points

Attacks can be identified in different areas of the Infrastructure architecture, and in each of these areas the available logs can be used. For example logs originating from firewall, IDS, DBMS and application firewalls (WAF) can be used to complete the knowledge about an attack.

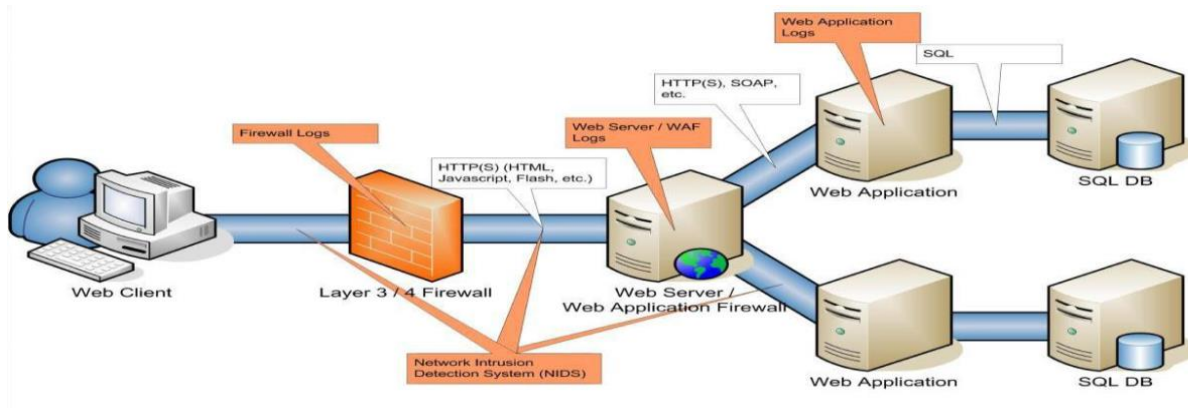

.

**Figure 21:** Infrastructure Architecture Control Points

All these rules can only be written if the attack vectors are known before writing down the rules. In this way, we can identify the attacks considered by OWASP to be among the 10 most critical attacks for the applications exposed on the Internet [2]. On the contrary, an approach to identifying statistical anomalies does not require prior knowledge of attack vectors. The nature of the work carried out anyway is experimental and indicative of the potential of the Splunk tool under investigation. It cannot be considered in terms of a product ready for the production environment.

# 8. PERFORMANCE CONSIDERATIONS

## 8.1 Further Performance Considerations on Search Queries

All the searches carried out in this project work have been found to have search times limited to 20-30 seconds. Excepted for the calculation of the load, the search spent 101.36 seconds, compared to 744,649 events consulted. It can be asserted that the execution times depend on the complexity of the search query, the total data to consult, the resources available on the machine (number of CPUs, RAM size, type of secondary memory) and also on the configuration of the Splunk platform. In order to experience the real dependence of execution times on HW resources the RAM size has been expanded, and the number of Virtual CPUs on the cloud has been increased, creating two new indexes: Indexer and IndexerLarge (Figure 22):

| NOME SERVER | IP PUBBLICO | HYPERV. | SO | VCPU | RAM | HD |
|---|---|---|---|---|---|---|
| UniversalForwarder SMART | 80.211.155.253 0% traffico usato | vmware | CentOS 7.x 64bit | 1 | 1 GB | 20 GB |
| Indexer SMART | 94.177.160.159 0% traffico usato | vmware | CentOS 7.x 64bit | 2 | 4 GB | 80 GB |
| IndexerLarge SMART | 188.213.174.205 0% traffico usato | vmware | CentOS 7.x 64bit | 4 | 8 GB | 160 GB |

**Figure 22:** Installed Virtual Machines

After reinstalling on Indexer and IndexerLarge two new configurations of data and programs, it was continued the work consisted on running search queries on the application load made previously. They were obtained the following times results (Figure 23).

**Ispezione processi di ricerca**

This search has completed and has returned 3.270 risultati by scanning 744.649 eventi in 55,052 seconds
(SID: 1526554014.1157) search.log

**Costi di esecuzione**

| Durata (secondi) | Componente | Chiamate | Conteggio input | Conteggio output |
|---|---|---|---|---|
| 0,05 | command.fields | 72 | 744.649 | 744.649 |
| 0,25 | command.pretransaction | 144 | 2.233.947 | 2.233.947 |
| 54,33 | command.search | 72 | - | 744.649 |
| 0,61 | command.search.index | 72 | - | - |
| 0,55 | command.search.fieldalias | 69 | 744.649 | 744.649 |
| 0,27 | command.search.calcfields | 69 | 744.649 | 744.649 |
| 0,04 | command.search.expand_search | 1 | - | - |
| 0,00 | command.search.index.usec_1_8 | 4 | - | - |
| 0,00 | command.search.index.usec_512_4096 | 2 | - | - |
| 0,00 | command.search.index.usec_64_512 | 60 | - | - |
| 0,00 | command.search.index.usec_8_64 | 122 | - | - |
| 35,05 | command.search.kv | 69 | - | - |
| 8,58 | command.search.typer | 69 | 744.649 | 744.649 |
| 6,67 | command.search.rawdata | 69 | - | - |
| 0,22 | command.search.lookups | 69 | 744.649 | 744.649 |
| 0,14 | command.search.tags | 69 | 744.649 | 744.649 |
| 0,02 | command.search.summary | 72 | - | - |
| 0,00 | command.search.parse_directives | 1 | - | - |
| 0,08 | command.timechart | 74 | 745 | - |
| 0,07 | command.timechart.execute_input | 73 | 745 | - |
| 0,00 | command.timechart.execute_output | 1 | - | - |

**Figure 23:** Process Inspection on the Indexer Virtual Machine

What emerges from this experimentation is that by doubling and quadrupling the power of the machine, i.e. going from 2GB of RAM to 4GB and then to 8GB of RAM, and also by doubling the virtual CPUs from 1 to 2, and finally to 4 CPUs, the execution time of the search query falls significantly. In the Figure 23 we can see the data resulting from the execution of the 'application load' query on the Indexer machine, (4GBs of Ram, 2 VCPUs).

In the following Figure 24 we can see the data resulting from the overall execution of the 'application load' query on the IndexerLarge machine (8GBs of Ram, 4 VCPUs).

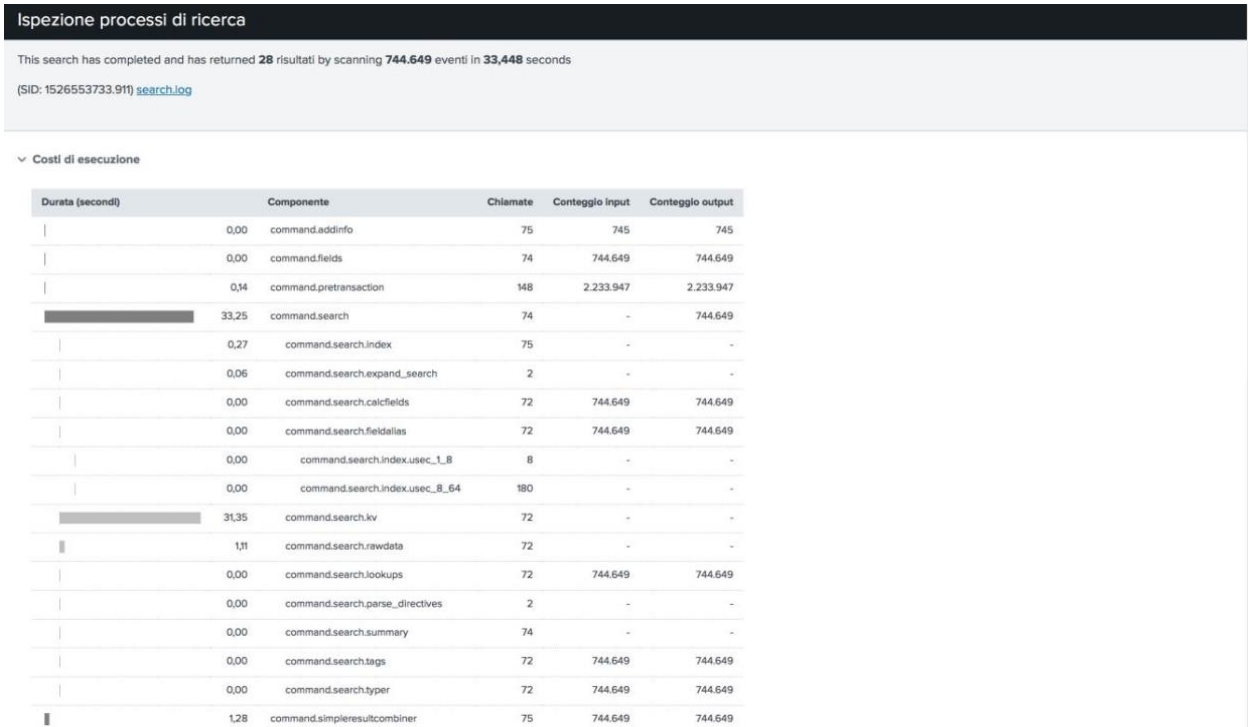

**Figure 24:** Process Inspection on IndexerLarge Virtual Machine

## 8.2 Costs of query execution

Therefore the query which lasted 101.36 seconds on the original virtual machine, has dropped to 55.052 sec. on the doubled power virtual machine, and then to 33.448 sec. on the quadruplicated power one. The law is almost linear so it can be assumed that with 32 GB of RAM and 16 CPUs the time spent in the search query will be about 7 seconds, or 3 seconds if a 64 GB and 32 CPUs will be used. This is in line with the HW requirements recommended by Splunk for Enterprise platforms.

From the process inspection above you can also see that the components that contribute most to the overall time of the queries are related to "command.search", in particular its component 'command.search.kv'.

In the case of execution with the major HW configuration, what happens is that after Splunk software identifies the events containing the indexed fields that meet the search queries, the events are further analyzed to see which of them also meet the other search criteria of the query.

The subprocess 'command.search.kv' (32.11 sec.) tells us how long it took to extract the fields from the events.

In this case it is more than 90% of the total time of the search and much more relevant than the component 'command.search.rawdata' (1.18 sec.) which indicates how long it takes to extract events from the file rawdata.

This aspect will be further analyzed in detail to understand why this happens, as behind these services may be hidden inefficiencies caused by the data.

## 8.3 Organization of Indexes in Splunk

Splunk index data are organized in **buckets**, which are directories.

There are 5 types of buckets: **hot, warm, cold, frozen** and **thawed**. Therefore the data in Splunk have a life cycle moving from hot, to warm, to cold and finally to frozen, making an automatic rolling as you reach the saturation dimension of the previous level. The hot type of bucket is the only one that is not necessarily optimized. Each bucket contains events related to time periods. The size of each of them can be decided by the user when creating and configured the index.

The search for a certain value, a field, always starts from the hot index, and when finished the search proceeds to the next "coldest" field.

When you search for a string and you find it in a bucket you must then access the rawdata to verify the event that contains the string you are looking for, but to do this must decompress the rawdata, which are stored in compressed format.

The more events that satisfy the search, the more decompressions will be necessary to do, and therefore the greater will be the commitment of the CPU.

If the decompressions are few, the CPU used to access the rawdata is a little amount of time. Therefore, this is our case in Figure 24, as the evidence of the process shows that since there are only 28 results (this is the value at the time of execution) out of a total of 744,649 scanned events, the search in question can be considered to belong to the category "scatter", according to the following classification:

- "**Dense"** search 1 result for every 1-1,000 indexed events
- "**Scatter**" search 1 result per 1,000-1,000,000 indexed events
- "**Rare**" search 1 result per 1,000,000-1,000,000,000
- "**Needle in a haystack**" search 1 result per 1,000,000,000-finished

## 8.4 Reasons for being cautious in the search

In the case of the search query under consideration, we want to further investigate the time "command.search.kv" used to extract the fields of the events, as in the search query there is no field, except the host metadata, and therefore it is inexplicable the high cost of time spent in this process.

The explanation, however, is that the default configuration of Splunk is such that all fields are always extracted. For this reason we set KV_MODE = none in the props.conf configuration field to disable the automatic extraction of key value pairs. Repeating the query execution will result in a 10% shorter time.

A further reason for attention is the high value of time spent in the subprocess "dispatch.fetch.rcp.phase_0" which is the time spent waiting for the fetching of events from the search.

The fetch is mostly waiting for events to be extracted from the disk. It may be caused by I/O operations. In fact, since it has been said that the process is not CPU bound, consequently the I/O component prevails over the processing component. A concause in this case could be a bad structuring of the data in the buckets, but in our case this hypothesis is to be discarded since the hot type bucket has not yet been saturated. Another cause could be the I/O disputes, in which case the reason could be attributed to the supplier of the VPS (Aruba provider), and therefore out of our control.

## 8.5 Other variables that impact execution times

Finally, let's consider other elements on which the execution costs could depend.

The size of the buckets is important, although it is usually self-contained and depends on the machine cores. Another element may be the "Bloom Filter". These are data structures that categorically establish that an element is not contained in a data set. In this way you can say, without opening a data set within which we are looking for a string, that a certain data is not contained in it.

This increases performance considerably. Bloom filters are created in the indexing phase and are active by default in Splunk .

The speed of writing and reading to disk also has an impact on search performance. The indexing phase depends on the write performance of the disks; while the search phase is partially dependent on the read performance. The SSD disk type performs best for searching and is estimated a logarithmic distance with the performance of the disks of mechanical type.

## 9. MACHINE LEARNING AND STATISTICAL APPROACH

### 9.1 Machine Learning to determine the anomalies

According to Gartner's PPDR (Predict, Prevent, Detect, Respond) model, all security tasks can be divided into five categories: **prediction; prevention; detection; response and monitoring**. We are focusing on the **detection** task of cybersecurity.

The methodology followed so far by using the Splunk tool was based on the identification of events made through regular expressions, by which the individual important fields where indentified.. On these fields I the queries were built and the results to be displayed in the dashboards were extracted. The dashboards were successively made dynamic and further explorable thanks to drill down.

Let's suppose that something could escape this process of log mining. The process we are going to describe is called 'Data Mining'. The process involves the extraction of patterns from the data and the fitting of the model built on it. The concept behind the fitting of the model is handle the information which can be inferred from processing such a model. That's why we evaluated the use of **Machine Learning** techniques, which allow us to discover those events that may have escaped our investigation or that were not thought to exist. Machine Learning is the machine's ability to learn automatically without being programmed by a man. It is a method to generalize from the data contained in the logs, and build models. The activities to be carried out in the Machine Learning activity are the following:

1. Collect data
2. Clean and transform data
3. Explore and view data
4. Modeling data
5. Evaluate and Refine the results of the model
6. Release the model (for future predictions and uses)

The activity carried out in the Machine Learning and statistical approach of this project work has consisted in the installation of the tool **MLTK vers. 3.2.** This is an app downloadable from Splunkbase and executable in the infrastructure, with which, thanks to the libraries of Python's scientific computing, the SPL commands can be extended to the field of Machine Learning. The MLTK toolkit allows to perform various activities including:

1. Predict numeric fields
2. Predict categorical fields
3. Identify numerical outliers
4. Identify categorical outliers
5. Provide time series
6. Cluster numerical events

Within this paper some experimentations applied to the application logs will be reported, and thus producing the results related to the activities 2. (prediction categorical fields) and 4. (identify categorical outliers). An anomaly is defined as the unusual behavior or pattern of the data. It particularly may indicate the presence of the error in the system and itdescribes that the actual result is different from the obtained result thus concluding that the applied model does not fit into the given assumptions. In other words the model drawn in the data mining phase indicates that the actual result is different from the obtained result from the model. In a more general meaning the **anomaly does not necessarily means an error** but **it indicates an highly probability of it.**

Both experiments play an important role in the determination of anomalies for the reasons mentioned at the beginning. In the present section, the results of Machine Learning activities will prove to be important both for the determination of malfunctions and also for the identification of events potentially subject to further investigation. By analyzing the log data anomalies the security of the system could be improved and the potential intrusions could be detected.

In the first part of this paper we have presented the reader some empiricals ways of detecting web attacks. The Machine Learning (ML) for Application Security is an area in which the use of Machine Learning can improve app security. There are web applications, or web services, or micro services, in which ML can solve some of tasks relevant to cybersecurity.

**9.1.2 Classification and Regression**

In order to apply the concepts of Machine Learning for application security there are two main possible ways :

- classification to detect known types of attacks like injections
- regression to detect anomalies in HTTP requests

These two techniques are applicable in different kind of problems. The former when we are looking for a classifications output (the output variable is over or below a given threshold ? is or is not ? etc. ). The latter is characteristic of a numerical and precise value of the answer (for example the output variable is 5 ? ). In the case under examination we will take into consideration a classification task, namely a prediction of categorical fields (activity 2).

But we can also distinguish regression from classification according a different aspect.

**Regression** (or prediction) is a task of predicting the next value based on the previous values.

**Classification** is a task of separating things into different categories.

### 9.1.3 Supervised and Unsupervised Learning

The tool at our disposal (MLTK) allows us to use both 'supervised' or 'unsupervised' learning.

That is, the construction of the model that is obtained is in the first case guided by examples (supervised) in which the system is made aware of the correspondence, in a set of training data, between the output (the field to be predicted) and the input (the so-called 'predictor' fields). In the second case it is assumed that the correspondence between the data of the input training set and those of the output is not known, and the same are treated indistinctly at the end of the construction of the model.

Logistic Regression, SVM, PCA are examples of supervised learning.

Linear Regression, KBMeans, DBSCAN, Spectral Clustering, are examples of unsupervised learning.

In both cases input data are provided to the Machine Learning algorithm, which will proceed to a training and testing phase of the model.

### 9.1.4 Data pre-processing and data transformation

In the work done in this project work the data are provided as a result of a search, coming from a CSV data source, and they have been previously 'cleaned' and then 'transformed'. The cleaning process consists in making the data as structured as possible and this takes place also by eliminating unstructured information that may occur within the logs.

Subsequently, the transformation can intervene, optionally, in the 'pre-processing' part of the data. In this phase the values can be 'standardized' with respect to the average and standard deviation, in the case of large numerical values that otherwise could complicate the calculations.

In addition, it may be necessary to bring the data back into tabular form, because often the scientific calculation algorithms , by which the data are processed, require matrices.

Pre-processing uses algorithms such as Fieldselector (which selects the fields that are the best predictors), PCA (which reduces the number of fields), KernelPCA (which reduces the number of sizes), StandardScaler (which standardizes the fields).

In this work reported in this paper we have carried out pre-processing, using the KernelPCA algorithm, in activity 2. (prediction of categorical fields), in order to reduce the number of fields, i.e reduce the size, and make more efficient the computation. In some cases KernelPCA or PCA can be used, to make it easier to view the final data, reducing the size to two.

### 9.2 Category Outlier Prediction

### 9.2.1 Category Outlier Prediction

Let's see below the phase of the pre-processing (Figure 25):

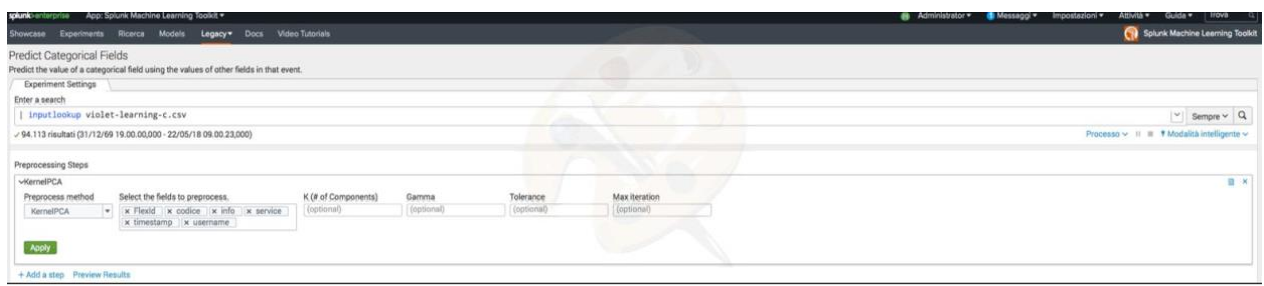

**Figure 25:** Category Outlier Prediction Preprocessing

After setting the method , the individual fields to be pre-processed and other optional fields, it must be pressed the 'Apply' button to start the pre-processing.

At the end of the operation the data will be visible in a preview with the pre-processed format, including fields PC_1, PC_2, related to the resizing.

It is also possible to put in sequence other pre-processing phases, adding a step, with the 'Add a step' key.

Once the pre-processing is finished, we will have the data in the desired format, and we will be able to pass to the real 'learning' phase.

### 9.2.2 Learning phase :'fit' and 'apply'

The MLTK tool allows the user to extend the SPL language, which is thus enriched with new search commands, among which the most used are :

- **'fit'** is a command that serves to 'learn' a model from the data, search results (or lookup)

- **'apply'** is a command that applies a learned model (with the fit command) to a new data set

The "fit" command requires you to specify the algorithm to be used.

Some of these algorithms are: LogisticRegression, LinearRegression, OneclassSVM , BernoulliNB, GaussianNB, GradientBoostingClassifier, RandomForecastClassier, SGDClassifier, SVM.

Will not go into the details of the algorithms, which belong to the open source libraries of Python (more than 300 open source programs),  of the libraries sci-kit learn, pandas, statsmodel, numpy and scipy, available thanks to the add-on Splunk Python for scientific computing.

The result of the 'fit' command will be the model, which  will be given a name (see in the Figure 26, the name given is 'example_prova') , and which will be an object of the Splunk platform, and therefore reusable in a later moment, with different data sets.

The 'fit' phase of the model for the previously pre-processed data will appear as shown in the following Figure 26:

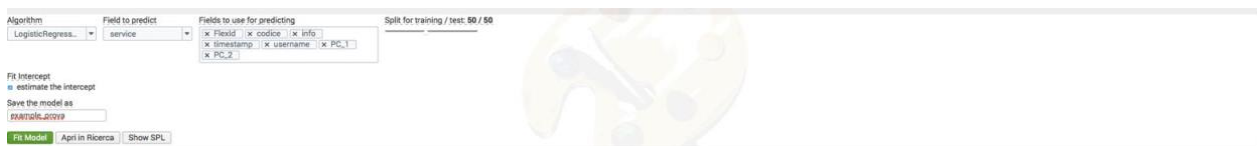

**Figure 26:** Model Fit Phase

The Figure 26 shows that the Logistic Regression algorithm will be used as a supervised learning method, in which the data for training and testing are 50% each.

Model validation consists in training the model with a portion of the data (training set) and testing the model with another portion (test set).

Moreover, the field to be predicted, called response field, is 'service'. The 'service' field is a field of the log, which has been appropriately pre-processed, and which indicates the application service used.

The remaining fields are predictor or explanatory fields (i.e. info, timestamp, FlexId, code). Each activity of the learning can be implemented using different algorithms such as linear regression, logistic regression, statistical distribution, probability measurements, Kalman Filtering states space method, K-Means, spectral clustering, Birch, DBSCAN.

The model, including the pre-processing phases, is stored with its identifier and can then be used to 'apply' it to another set of data.

The commands of the SPL language to realize the 'fit' phase of the model are the following:

"| inputlookup indexB-learning.csv | FIT KernelPCA "FlexId", "codice", "info", "service", "timestamp", "username" k=2 INTO modello_previsione_KernelPCA_0 | FIT LogisticRegression fit_intercept=true "service" FROM "FlexId" "codice" "info" "timestamp" "username" "PC_1" "PC_2" INTO "modello_previsione" "

The model so far created can subsequently be stored among Splunk's objects, with a view to its possible reuse. The next action is to apply the model to the remaining part of the data. The following SPL command will be issued to splunk :

"| inputlookup indexB-learning-c.csv | apply modello_previsione_KernelPCA_0 | apply "modello_previsione" | classificationstatistics("service", "predicted(service)" "

These commands apply pre-processing models and then apply the prediction model to the entire data set to predict the "service" field. The macro 'classificationstatistics' is also used to calculate the precision, recall, accuracy, and F1, which are quantities that indicate the goodness of the model obtained. The closer these values are to 1, the best the model is considered.

At the end of the execution of the apply commands the following table will be displayed (Figure 27):

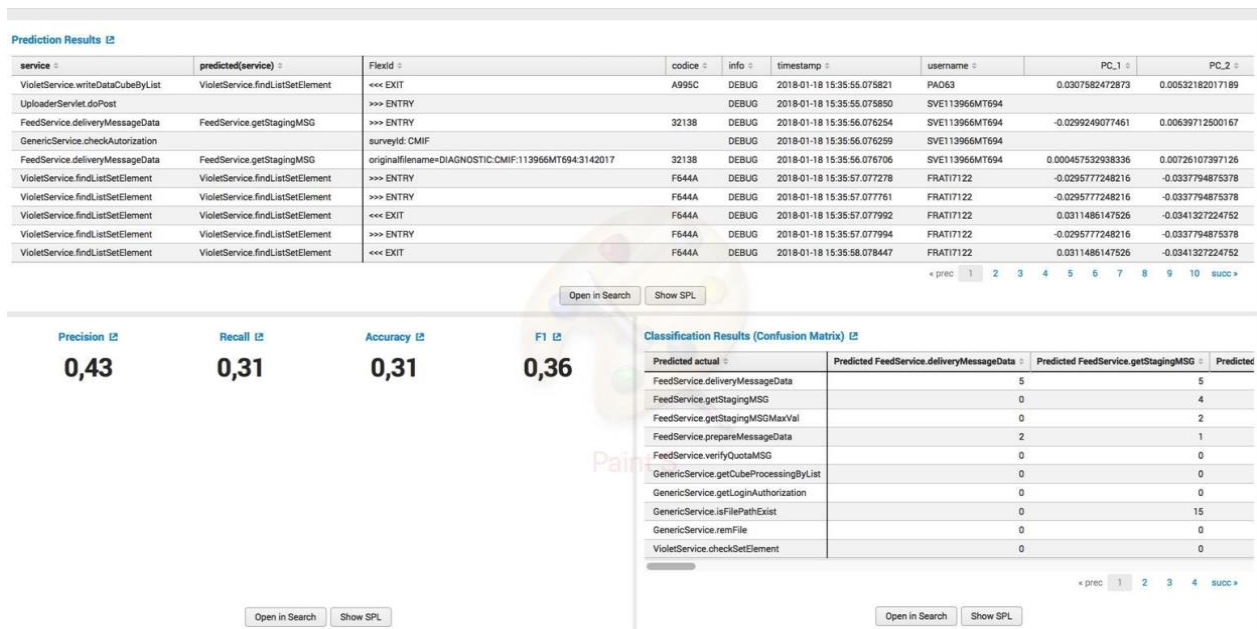

**Figure 27:** Result of applying the data model

A Confusion Matrix will also be created that will indicate the result of differences between the predictive model compared to real data. The more the data will be distributed along the diagonal, the more accurate the model will be (Figure 28).

**Figure 28**: Confusion Matrix

### 9.2.3 Final result of model processing: refining of the model

The final result that will be obtained is the result of successive attempts in which all the parameters have been adjusted in order to obtain a result as reliable and accurate as possible. The process of adjusting the parameters is called 'refining of the model'. This is the hardest and most expensive part of the work, also because it requires resources and time.

A model can be refined by removing distracting fields or even adding more fields to increase accuracy.

Once the model has been validated and refined can be considered valid and therefore usable in all those cases of log data having the same structure.

The meaning that can be drawn from the experimentation reported so far is that the anomaly becomes evident when there is a deviation between the current result and that predicted, with the premise that the model is considered a **good model**. The anomaly must therefore be investigated and further elaborated as a sign of a 'potential' problem.

We tried to repeat the experiment carried out with respect to the field to be predicted 'service', also with respect to the field 'username', obviously with other predictors, and we obtained an even more reliable model, with a precision = 0.99.

The result of this elaboration is shown below (Figure 29) :

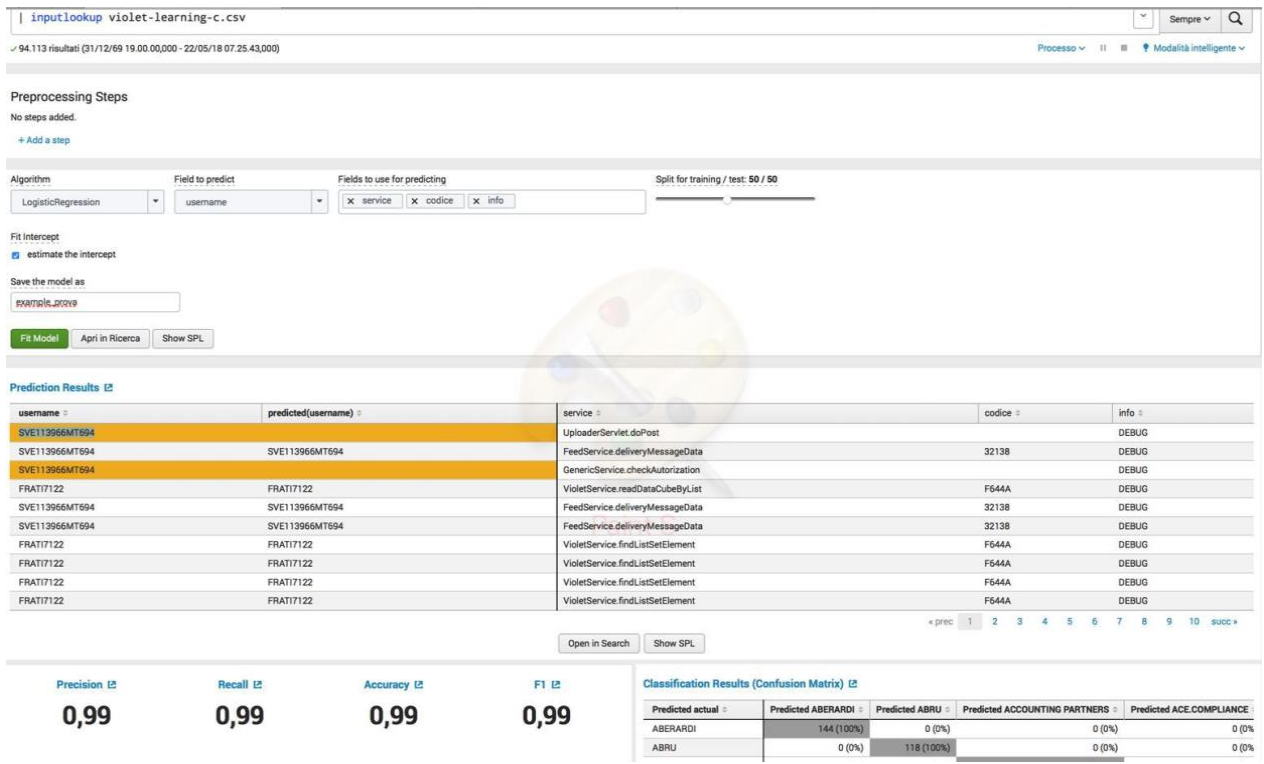

**Figure 29:** Output of the prediction model

### 9.2.4 The suggestion for further investigations

The approach outlined here is to be understood as a starting point for subsequent investigations and certainly not a deterministic and exhaustive approach. For this reason, it is a useful tool in the early stages of forensic investigations related to logs, when you do not yet know where to direct attention or when, even after analyzing everything, you still want further evidence that you have not missed anything. What must be said is that predicting categorical fields is a type of learning known as classification. This type of algorithm learns the tendency of data to belong to one category or another, based on the related data. Therefore what emerges from the visualization of the above tables is the actual state of the username field, versus the expected state of the same field. It should be noticed, in yellow, the incorrect prediction from the model, and therefore the anomaly. Although it has not been treated in the elaboration here reproduced, a further possible application of this technique, could be the forecast of the application load, according to other applicative variables. This experiment, which presupposes a complex preparation of the data, would allow the provisioning of the resources in cases of high load forecasting, and therefore would have repercussions on the management costs.

### 9.3 Categorical Outlier Identification

In this paragraph the experimentation related to the identification of the categorical outliers (activity 4.) is reported. This activity it is not to be considered a modeling of data. The input file is, as in the previous task, a CSV data source, and it must be cleaned through a process analogous to the previous section by making the data as structured as possible, made by eliminating unstructured information that occurs within the logs. After using the SPL command 'anomalydetection' with which the outliers could be determined.

**"| inputlookup indexB-learning-c.csv | anomalydetection "service" "timestamp" "username" action=annotate | eval isOutlier = if(probable_cause ¯"", "1", "0") | table "service" "timestamp" "username", probable_cause, isOutlier | sort 100000 probable_cause"**

This command is not really a 'Machine Learning' command but applies **probabilistic techniques** to events by identifying anomalous events and calculating the probability of each event, and then identifying the smallest probabilities. Probability is the result of the frequencies of each value in the individual event fields. In the process of identifying anomalies of the data in the logs, we can proceed with a 'univariate' or with a 'multivariate' approach. That is, in order to identify anomalies we can determine the individual anomalies for each field, taken individually; otherwise we can execute the command to identify the anomalies by specifying at the same time several fields in relation to which we want to determine the anomalies. It is obvious that the results obtained with the '**univariate**' approach do not coincide with the '**multivariate**' one, because the method of calculating the probability is different. In the former case it is a probability of a simple event, in the latter case it is a compound probability. The determination of the outlier values can be used as the trigger of an alert or a subsequent action. Below is reported the experimentation on the determination of the categorical outliers (anomalies) made on the single field 'service' (Figure 30):

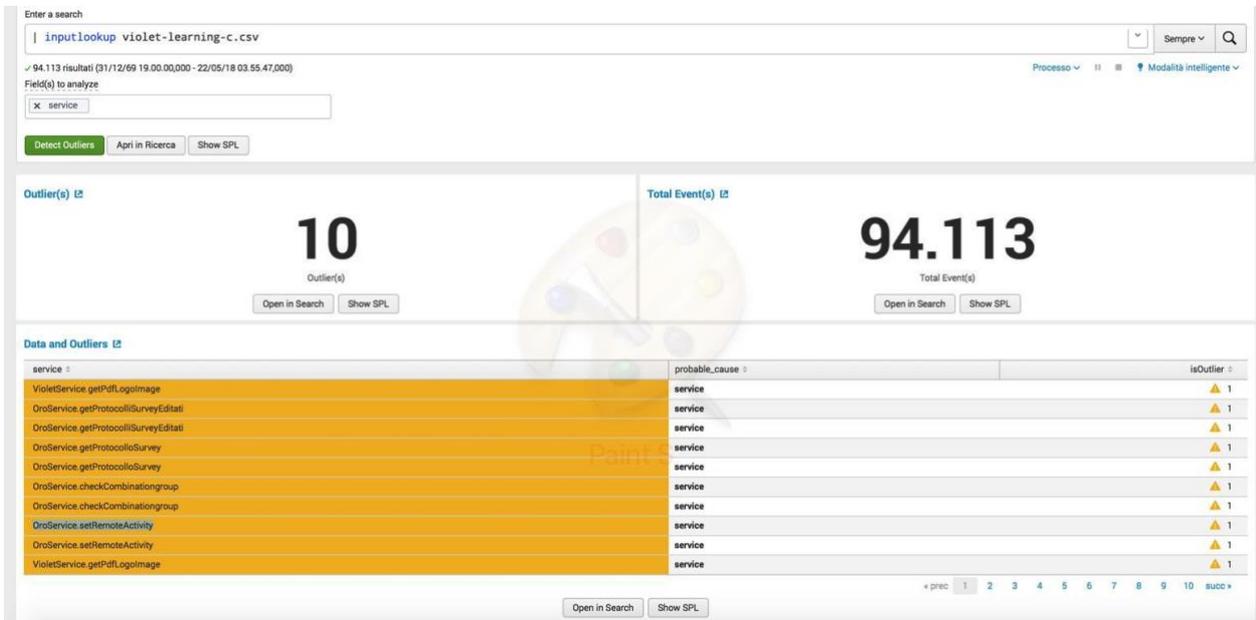

**Figure 30:** Unique categorical output on 'service' field

In the following Figure 31, it is reported the determination of the outliers related to the single field 'username':

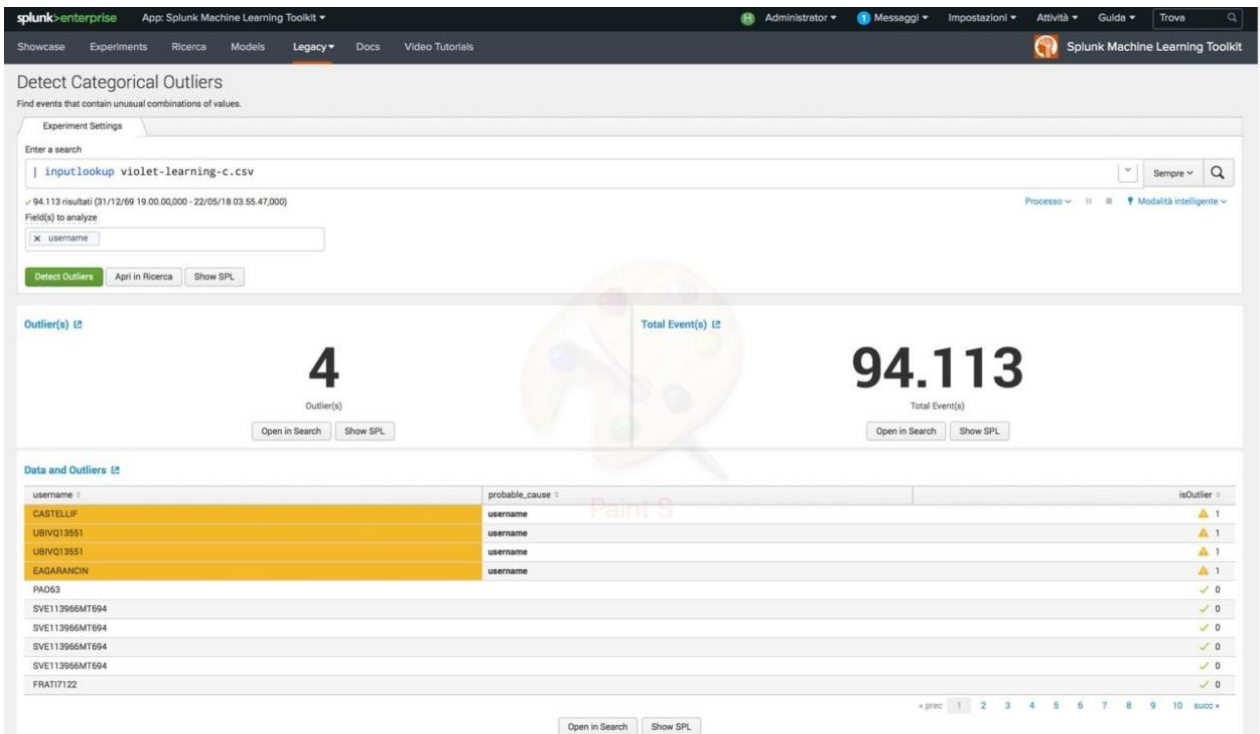

**Figure 31:** Unique categorical output on field 'username' field

In the following Figure 32, the reader can observe that, if they are added multiple fileds , i.e. 'username', 'service' and 'timestamp', the result obtained as output table is a different one, thus indicates different outliers from the previous ones.

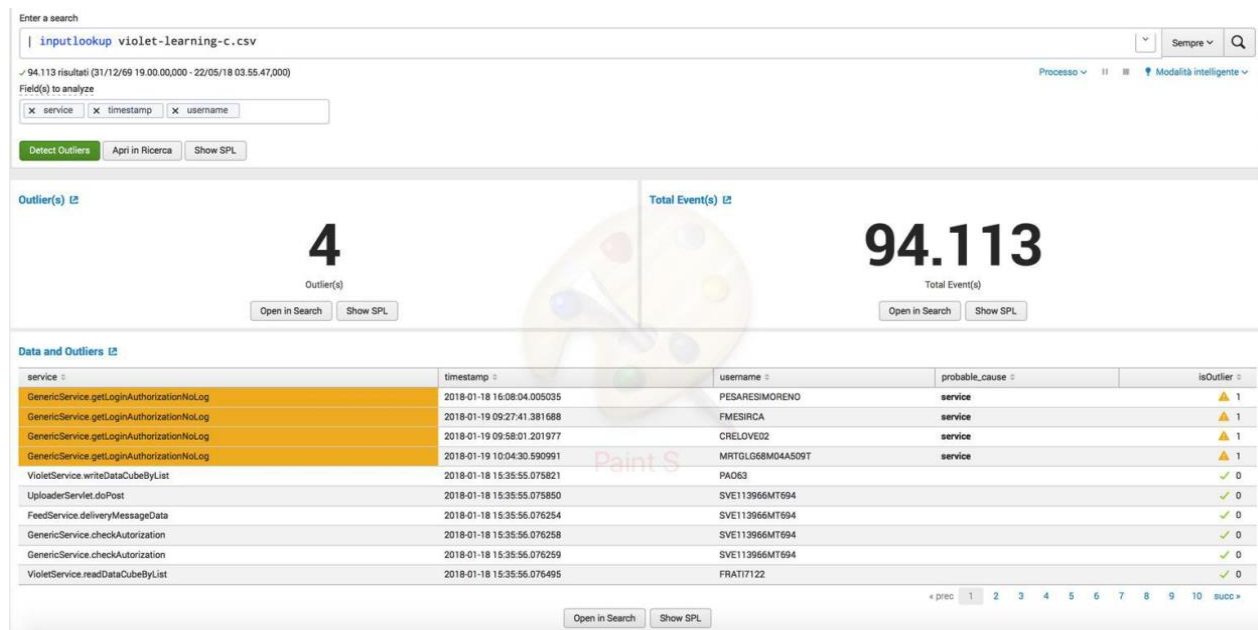

**Figure 32:** Multivariate categorical output

Note that the results showing the outliers are accompanied by visual information on their number, on the overall events, on the reason why those events are considered outliers. This report can be a starting point to investigate furtherly the anomalies, and to verify if there are any fraudulent behaviors behind them. In the image highlighted in yellow the data indicating unusual events.

## 10. CONCLUSIONS

My work begun with the aim of enhancing the application log files, thus making them company assets. An asset implies adequate protection mechanisms, professional skills that deal with it, adequate training, standards of writing and compliance. In this work can be assessed how much value can be extracted at an informative level from the analysis of the logs, exclusively considering those logs of applicative nature. Backwards, the work retraced the reverse path of the synthesis of the data analyzed, to make them visible and usable from the perspective of Business Analytics. It was observed if it is more useful in the process of log analysis to "look at the tree or the forest", and thanks to a tool as powerful as versatile and scalable as Splunk it was concluded that both could be done. Forensic analysis is no longer an investigative activity at Sherlock Holmes way of doing, that is, using a magnifying glass to go deeper and deeper into the details that can trace the anomalous event or fraudulent compromise, but it must be considered an overview. It also have been examined the performance aspects of the architecture, capacity planning and dimensioning from an enterprise point of view, since the future lies in Big Data, that is, in the processing of large amounts of data, in real time, and with data variable formats and structures. Finally it was seen how intelligence can be added to the discovery process thanks to Machine Learning and to the probabilistic approach. The next steps will consist in increasing the ability to correlate logs of different origins, aggregating events that originate from the same action, reasoning on time, synchronizing events that may present apparent misalignments, and involving multiple devices. This aspect is all the more important since APT (Advanced Persistence Threats) attacks are going to spread. These attacks are characterized by being prolonged, persistent, and wide-ranging on multiple targets of the information system. A limit of the present work at the moment is that the analysis is based only on the information contained in the logs and not on all other sources such as online traffic data, scripts execution, API calls, commands, which would provide a certain vision even more complete, but at the moment outside the range of action of the presented project.

ACKNOWLEDGEMENTS

Special thanks should be given to Dr. M. Bernaschi, Dirigente Tecnologo" of the National Research Council of Italy and also Adjunct Professor of Computer Science at Rome University "Sapienza", my research project supervisor for his professional guidance and valuable support , his useful and constructive recommendations on this project.

REFERENCES

[1]   Towards structured log analysis, by Dileepa Jayathilake, from International Joint Conference on Computer Science and Software Engineering (JCSSE), Publication Year: 2012, Page(s): 259 – 264

[2]   A. Krishna, Splunk Admin & Architect: Complete Tutorials + 30 Days Lab, Udemy, Online.

[3]   Exploring Splunk search processing language (SPL) primer and cookbook, by David Carasso, Splunk's Chief Mind, CITO Research New York, NY , 2012

[4]   R. Meyer, Detecting Attacks on Web Applications from Log Files, SANS Institute , 2008, Roger Meyer

[5]   Splunk and the SANS Top 20 Critical Security Controls, Mapping Splunk Software to the SANS Top 20 CSC Version 4.1, 2014 by Splunk Inc.

[6]   Splunk Inc, Splunk Machine Learning Toolkit User Guide 3.1.0. 2018

**Authors**

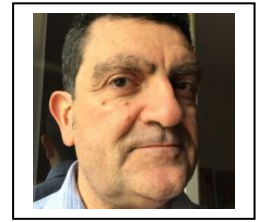

Short Biography

Dr. Roberto Bruzzese is an employee of a private institute under Italian public law. He graduated in Computer Science from the University of Bari in 1986. He won a scholarship for scientific research in 1988 and worked for three years as a researcher at Tecnopolis Csata N.O., Valenzano, Italy (1990). Since 1991 he has been working as a developer in the Department of Informatics, of a private institute under Italian public law in Rome, Italy. He subsequently obtained his specialization in Web Technologies at the University of L'Aquila (2015), and a second specialization in Cybersecurity at the University of Rome, La Sapienza.  His current fields of study are Internet of Things, Malware Analysis, Ethical Hacking.